\documentclass[aps,pre,onecolumn,longbibliography,showpacs,amsmath,amsmath,amssymb,amsfonts,superscriptaddress,notitlepage]{revtex4-1}
\usepackage{graphicx}
\usepackage{dcolumn}
\usepackage{bm}
\usepackage{color}
\usepackage{multirow}
\usepackage[hypertexnames=false]{hyperref}
\usepackage{cancel}
\usepackage{natbib}
\begin{document}
\title{Steering complex networks toward desired dynamics}

\author{Ricardo Guti{\'e}rrez}
\email[Correspondence should be addressed to: ]{rigutier@math.uc3m.es}
\affiliation{Complex Systems Interdisciplinary Group (GISC), Department of Mathematics, Universidad Carlos III de Madrid, 28911 Legan{\'e}s, Madrid, Spain}
\author{Massimo Materassi}
\affiliation{CNR, Institute of Complex Systems, Via Madonna del Piano 10, 50019 Florence, Italy}

\author{Stefano Focardi}
\affiliation{CNR, Institute of Complex Systems, Via Madonna del Piano 10, 50019 Florence, Italy}

\author{Stefano Boccaletti}
\affiliation{CNR, Institute of Complex Systems, Via Madonna del Piano 10, 50019 Florence, Italy}
\affiliation{Unmanned Systems Research Institute, Northwestern Polytechnical University, Xi'an 710072, China}
\affiliation{Moscow Institute of Physics and Technology (National Research University), 9 Institutskiy per., Dolgoprudny, Moscow Region, 141701, Russian Federation}
\affiliation{Universidad Rey Juan Carlos, Calle Tulip\'an, s/n, 28933 M\'ostoles, Madrid, Spain}

\begin{abstract}
We consider networks of dynamical units that evolve in time according to different laws, and are coupled to each other in highly irregular ways. Studying how to steer the dynamics of such systems towards a desired evolution is of great practical interest in many areas of science, as well as providing insight into the interplay between network structure and dynamical behavior. We propose a pinning protocol for imposing specific dynamic evolutions compatible with the equations of motion on a networked system. The method does not impose any restrictions on the local dynamics, which may vary from node to node, nor on the interactions between nodes, which may adopt in principle any nonlinear mathematical form and be represented by weighted, directed or undirected links. We first explore our method on small synthetic networks of chaotic oscillators, which allows us to unveil a correlation between the ordered sequence of pinned nodes and their topological influence in the network. We then consider a 12-species trophic web network, which is a model of a mammalian food web. By pinning a relatively small number of species, one can make the system abandon its spontaneous evolution from its (typically uncontrolled) initial state towards a target dynamics, or periodically control it so as to make the populations evolve within stipulated bounds. The relevance of these findings for environment  management and conservation is discussed.
\end{abstract}

\maketitle

\section{Introduction}

Controlling the dynamics of ensembles of units networking via irregular topologies is one of the foremost challenges of modern science, and, in fact, the literature of the last two decades abounds with proposals for network control. In some of the earliest contributions to the field, a pinning  method based on applying linear feedback injections to some nodes of a network with the objective of stabilizing a given global fixed point was explored \cite{wang2002pinning, wang2002synchronization}.  Pinning controllability was further studied in Ref.~\cite{sorrentino2007controllability} as a way to synchronize to a given, time-dependent, network evolution. Similar approaches that expand or modify these initial efforts were developed in more recent contributions, see e.g. \cite{xiang2007pinning}. Later on, the introduction of multi-layer network representations  \cite{boccaletti2014structure} opened up new avenues, such as the study of complex-network targetability \cite{gutierrez2012targeting}, based on considering an identical copy of the graph undergoing a desirable evolution, and gradually creating unidirectional actions from nodes of the copy to the corresponding nodes in the original network, until the latter becomes fully synchronized with the former. These and other related works follow the master stability function approach \cite{pecora1998master} in assuming that dynamical units are identical, and that their coupling function at each link is the same, in order to derive analytical criteria for controllability.

A different approach was proposed in Ref.~\cite{liu2011controllability}, where conditions based on classical control and graph theories were given for the identification of the minimal set of nodes that, if forced to follow a prescribed time evolution, suffice to drive the entire network to the target dynamics. This is applicable to graphs whose dynamics is unknown, and for directed and weighted connectivities, when weights may possibly be unknown too.  This framework was also used to investigate network properties and their connection to structural controllability \cite{posfai2013effect, jia2013emergence,ruths2014control}. However, all these results come at the price of introducing drastic restrictions, as the study focuses on scalar state variables governed by linear equations of motion. In a recent contribution, moreover, control frameworks that are purely based on network topological properties (and completely ignore dynamical considerations), such as this one or the one proposed in Ref.~\cite{nacher2012dominating}, have been shown to fail in Boolean networks and models of biochemical regulation \cite{gates2016control}. Later developments along these lines include Ref.~\cite{wang2012optimizing}, which develops a perturbation approach to optimize the structural controllability of a complex network, and Ref.~\cite{yuan2013exact}, which generalizes the approach to a wider set of topologies via spectral techniques.  In one way or another, all such methods rely on rather restrictive sets of assumptions that are not always fulfilled in applications. This does not just mean that, at some level, there is always some degree of approximation: it may just be the case that different assumptions lead to radically different results. For instance, according to Ref.~\cite{liu2011controllability} the denser and more homogeneous a network is, the fewer nodes are needed to control its dynamics (though the conclusion appears to be different in later refinements aimed at an efficient choice of driver nodes \cite{gao2014target}), whereas the method proposed in Ref.~\cite{gutierrez2012targeting} comes to diametrically opposed conclusions. Other network control schemes that, strictly speaking, do not belong to any of the categories above have also appeared recently (see e.g. Refs.~\cite{cornelius2013realistic,zanudo2017structure}).

The most common assumption found in the network control literature is that the dynamical units are identical (which greatly simplifies both analytic and numerical treatments). Depending on the problem at hand, this may or may not be a drastic simplification: in physics it is sometimes a sensible approach, while it is clearly not in other disciplines, e.g.\! in the study of ecological systems. On the other hand, the assumption that the dynamics can be captured by linear ordinary differential equations is certainly not realistic for most applications, as it effectively bans limit-cycle or chaotic oscillations. Moreover, assuming linearly interacting units constitute also a severe limitation, as in most circumstances systems interact non-linearly: many-body gravitational and electrostatic problems in physics include, for instance, forces that are inversely proportional to the square distance between the interacting bodies, and a full treatment of solid-state and molecular systems frequently requires the incorporation of anharmonicities. In other areas of science nonlinear interactions are also the norm: in the modelling of ecosystems predator-prey couplings or competition for resources among species take the form of products of different populations, or more elaborated functional forms, see e.g. Ref.~\cite{holling1966functional}. While in some cases a linear (first-order) approximation might be justified, in some others it may even be not possible at all, as the coupling functions might not be analytic (as in models of neurons, whose action potential is fired when the membrane potential reaches a threshold). Lastly, another common (and quite drastic) assumption is that of having identical coupling functions represented by either directed (unidirectional case) or undirected (bidirectional case) networks, whereas  many systems (particularly in biology and social sciences) display in fact mixed couplings implying a combination of bidirectional and unidirectional links, with strengths and even functional forms that may vary from one link to another.

In this work, we introduce a technique for pinning control of networks that does not rely on any of these assumptions and is thus of wide applicability. The basic mechanism, previously introduced in a considerably more restrictive setting \cite{gutierrez2012targeting}, consists in establishing unidirectional pinning actions from a copy of the networked system (in practice it may be an experimental recording, or just a simulation of the dynamics) to the system on which one wants to impose the dynamics of the copy. In the jargon of multilayer networks, this is an inter-layer synchronization problem \cite{sevilla2016inter,leyva2017inter}: while individual nodes on a layer (the original network) may not be synchronized to each other, each of them is synchronized to its counterpart on the other layer (the copy). By considering synthetic mixed networks of nonlinearly coupled chaotic oscillators, we first derive some general results on the correlations between the nodes that need to be pinned and their topological properties. In essence, we find that those nodes that are influential on the dynamics of many other nodes but are simultaneously less influenceable by the rest of the network are by far the most efficient in setting inter-layer synchronization already with a small number of actions. This analysis also serves to illustrate the method in a relatively simple setting, yet including several features that violate the assumptions used in the  above-discussed references. 

We then illustrate the applicability of our method to real-world networks by steering the dynamics of a trophic web containing 12 species toward a desired evolution. This allows us to obtain information on which are the appropriate species to target, i.e.\! which species are keystone in the environment, as well as the best strategies to impose given dynamics on an ecosystem.  We discuss how these results can be used as a basis for adaptive management of ecosystems. Such a method can be  effective to foster the implementation of adaptive ecosystem management as requested by the application of Malawi principles of the Convention for Biological Diversity, \url{http://www.uni-kiel.de/ecology/users/fmueller/salzau2006/studentpages/Malawi_Principles/index.html}. From a formal point of view, this is a challenging networked system to control: its units (the species) are governed by different nonlinear equations, they are nonlinearly coupled via different coupling functions, and the pattern of connections is highly asymmetric and irregular (including different functional forms). This implies a strong departure from the set of assumptions used in all previous methodologies. After almost two decades of intense activity in the field, it is fair to say that none of the previous methods, as far as we are aware, can be applied to such a control problem despite its great environmental interest.

\section{Description of the method and application to networks of chaotic oscillators}

We consider a two-layer network. One layer is the {\it slave} layer, which corresponds to the original network over which one wants to impose the desired dynamics (i.e.\! a given evolution compatible with the equations of motion).  The other layer is the {\it master} layer, which is identical to the slave layer, but starts from a different initial condition (i.e.\! the one generating the specific desired dynamics towards which the state of the slave layer is to be steered), and evolves autonomously. In applications, the master layer may just be an experimental recording or a simulation of the original system ---as long as it can be coupled to the slave network its physical nature is irrelevant. Our control method consists then in establishing directed inter-layer links from nodes in the master layer to their counterparts in the slave layer. Once they are established, these links remain in place as more nodes are connected in sequential control steps. At each step the selected node is the one whose pinning causes the most rapid approach towards inter-layer synchronization (i.e. the imposition of the evolution followed by the master layer on the slave layer). While the two layers have to be identical, the nodes (i.e. the dynamical units) and links (the coupling structure  connecting the dynamical systems) on each layer can be completely different, as we will see below. This is thus a generalization of the method proposed in  Ref.~\cite{gutierrez2012targeting}.

We illustrate our method by applying it to networks of identical chaotic oscillators, and leave the applicability to more challenging real-world systems to the next section. Specifically, we consider networks of $N=50$ nodes whose topology is that of a mixed random graph, i.e. containing both bidirectional and unidirectional links. These graphs are realizations of the configuration model \cite{bender1978asymptotic} with the in-degree $k_\textrm{in}$ (i.e.\! the number of links pointing to a given node) and the out-degree $k_\textrm{out}$ (i.e.\! the number of links emanating from a given node) uniformly distributed in $\{5,6,\ldots,45\}$. Each node evolves autonomously in time as a chaotic R\"ossler oscillator, which we simply denote as $\dot{\bf r} = {\bf f}({\bf r})$, where ${\bf r} = (x,y,z)^\textrm{T}$ and $\dot{x} = -y - z,\ \dot{y} = x + a y,\ \dot{z} = b + z (x - c)$, with parameters $a=0.2$, $b=0.2$ and $c=7$. Nodes are coupled quadratically via their $z$ variables, a nonlinear coupling form that was previously considered in Ref.~\cite{liu2008phase}.

Before the first control step is applied (prior to the creation of the first inter-layer connection) both master and slave layers evolve spontaneously as follows
\begin{equation}
\dot{\bf r}_i =  {\bf f}({\bf r}_i) + \sigma_1 \sum_{j=1}^N D_{ji} (z_j^2-z_i^2) =  {\bf f}({\bf r}_i) + \sigma_1 \sum_{j=1}^N \mathcal{L}_{ji} z_j^2.
\label{fulleqs}
\end{equation}
where $D_{ji} = 1$ if there is a directed link from node $j$ to node $i$, and is zero otherwise (for bidirectional links $D_{ij} = D_{ji}$). As we do not consider self-links, the diagonal terms vanish, i.e. $D_{ii} = 0 \ \ \forall\, i$, and the in-degree of node $i$ is $k_{\textrm{in},i} = \sum_{j} D_{ji}$. The graph can thus be alternatively represented by the Laplacian matrix $\mathcal{L}_{ji} =D_{ji} - k_{\textrm{in},i} \delta_{ji}$.  The vector field ${\bf f}({\bf r}_i)$ governs the dynamics of node $i$, which would evolve autonomously (if uncoupled from its neighbors) simply as $\dot{\bf r}_i = {\bf f}({\bf r}_i)$,  and the parameter $\sigma_1$ is the intra-layer coupling strength

When the control procedure starts, each node $i$ in the master network keeps evolving according to the dynamics in Eq.~\ref{fulleqs}, $\dot{\bf r}^M_i =  {\bf f}({\bf r}^M_i) + \sigma_1 \sum_{j} \mathcal{L}_{ji} (z^M_j)^2$. In the slave layer dynamics, however, one has to consider an additional term which accounts for the inter-layer coupling from the master layer (without loss of generality, we here take a linear coupling through the $y$ variable). One has
\begin{equation}
\dot{\bf r}^S_i =  {\bf f}({\bf r}^S_i) + \sigma_1 \sum_{j} \mathcal{L}_{ji} (z^S_j)^2 + \sigma_2 \chi_i (y^M_i - y^S_i).
\end{equation}
Here $\chi_i$ is a binary variable that is one if there is a link coupling node $i$ in the master layer to node $i$ in the slave layer (i.e.\! if the targeting procedure includes a pinning action from master to slave at node $i$) and is zero otherwise. The parameter $\sigma_2$ is the inter-layer coupling strength. We emphasize that the coupling that is linear (in fact, diffusive) is the externally-imposed inter-layer coupling, which does not restrict in any way the form of the (intra-layer) couplings between the nodes of the system under study. Such diffusive inter-layer coupling is chosen as it is the simplest form that makes the inter-layer synchronization manifold into an invariant set of the dynamics (for a detailed mathematical treatment of invariant sets and related concepts, see e.g.\! Ref.~\cite{perko2013differential}).

\begin{figure}[t!]
\hspace{-0.5cm}\includegraphics[scale=0.44]{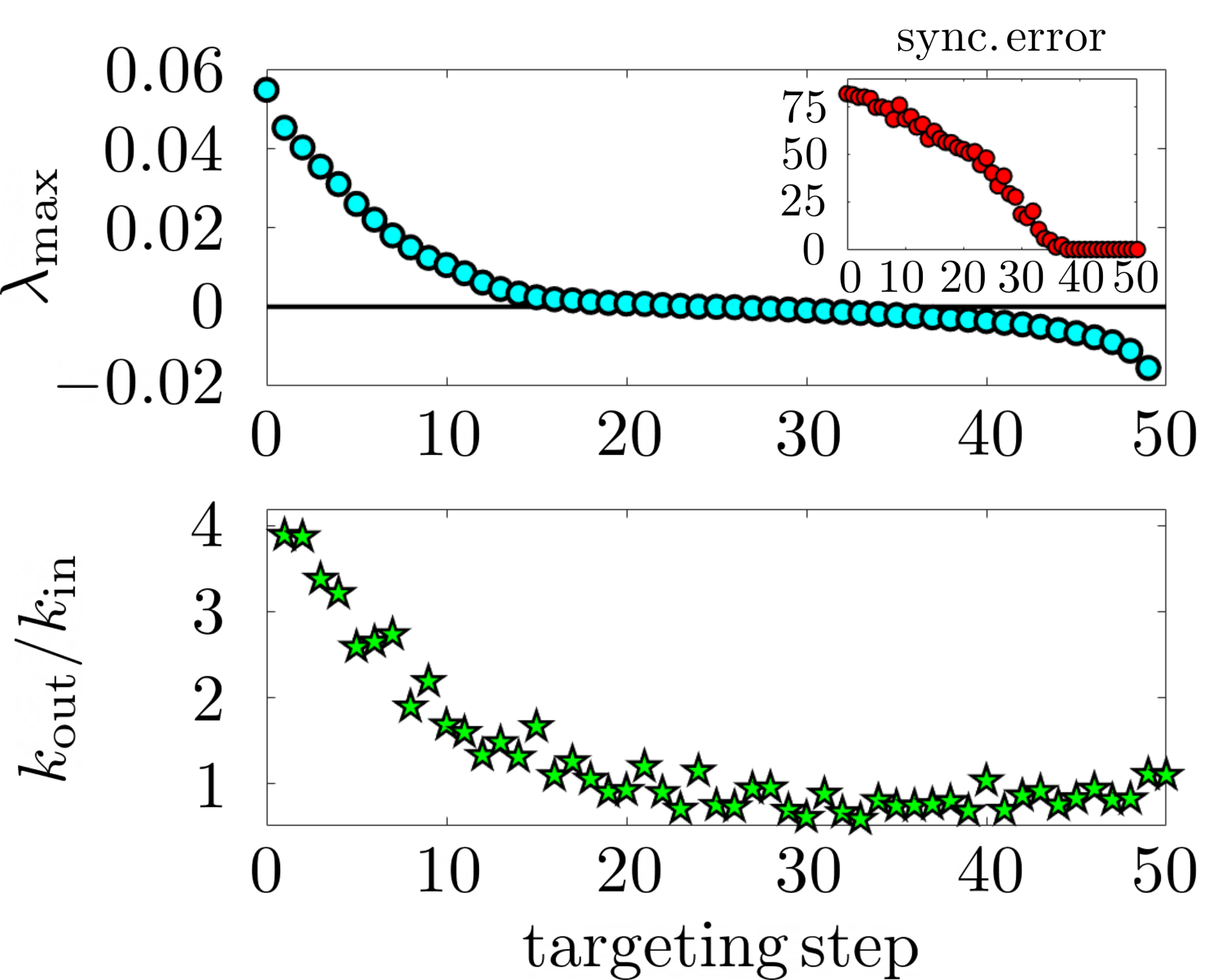}
\caption{ {\sf \bf Controlling the dynamics of a mixed network with uniform $k_\textrm{in}$ and $k_\textrm{out}$ distributions comprising $N=50$ nonlinearly-coupled R\"ossler oscillators with intra-layer coupling $\sigma_1 = 0.01$ and inter-layer coupling $\sigma_2 = 1$.}
(Top). Maximum Lyapunov exponent $\lambda_\text{max}$ (main panel) and synchronization error (inset) as functions of the targeting step.  (Bottom) Influence index $k_\text{out}/k_\text{in}$ of the node that is pinned at each targeting step. The curves are averages of 20 different network realizations.  A 4th-order Runge-Kutta method with a step of 0.01 time units has been employed for the numerical integration of the systems of $3 N=150$ ordinary differential equations corresponding to each layer.}\label{fig1}
\end{figure}

The results of applying our method to the network of R\"ossler chaotic oscillators are shown in Fig.~\ref{fig1}. Two observables are employed to characterize the inter-layer synchronization between master and slave as more and more inter-layer links are established in successive targeting steps. One is the maximum Lyapunov exponent, $\lambda_\text{max}$, computed from the dynamics of the slave network linearized around that of the master network as in Ref.~\cite{gutierrez2012targeting}. For a review of the theory and numerical computation of Lyapunov spectra, see e.g. Refs. \cite{benettin1980lyapunova,benettin1980lyapunovb}. The other observable is the synchronization error, which is the time average of the Euclidean distance in phase space $\mathbb{R}^{N m}$ ($N$ is the number of nodes, $m$ is the phase space dimensionality of the node dynamics ---in our case $N=50$, $m = 3$) between the full state of the master layer and that of the slave layer,  $ \lim_{T\to\infty} \frac{1}{T} \int_0^T \sqrt{ \sum_{i=1}^{N} (x^M_i(t) - x^S_i(t))^2 } dt$. In practice, $T$ is finite, but orders of magnitude larger than the characteristic timescales of oscillation (thus the numerical convergence to an asymptotic value is guaranteed). In the top panel of Fig.~\ref{fig1}, we show the maximum Lyapunov exponent $\lambda_\text{max}$ as a function of the targeting step, which is seen to progressively decrease as more and more nodes are pinned. Analogous results in terms of the synchronization error are reported in the inset, which shows how the synchronization error becomes zero when $\lambda_\text{max}$ becomes negative.

The maximum Lyapunov exponent is also used to identify the node to be targeted at each step: of all the nodes that remain unconnected to their counterparts in the other layer, the one that, when a master-slave connection is established, leads to the largest decrease in  $\lambda_\text{max}$ is targeted next. An exploration of possible correlations between the resulting targeting sequence (i.e.\! the ordered list of nodes that are targeted at successive steps) and local topological properties yields a remarkable correspondence between the targeting sequence position and the ranking of nodes in terms of their \textit{influence index}  $k_\text{out}/k_\text{in}$, as shown in the lower panel of Fig.~\ref{fig1}.  This index is large when a node has a privileged position for influencing other nodes, while receiving very little influence from the rest of the network. No such correlations are observed for connectivity indices that are insensitive to the directionality of connections, such as $(k_\textrm{in}+k_\textrm{out})/2$, while correlations only based on   $k_\textrm{out}$ or  $k_\textrm{in}$ give considerably poorer results that those shown in the figure.  Other measures of connection directionality that we have inspected, such as $(k_\textrm{out} -  k_\textrm{in})/(k_\textrm{out} +  k_\textrm{in})$, show weaker correlations with the targeting sequence than the influence index does. While these results are based on networks with uniform distributions of  $k_\textrm{in}$ and $k_\textrm{out}$, which have been chosen precisely because a large variety of possible degree values is desirable, a strong correlation between the influence-index ranking and the targeting ranking is also observed for Barab\'asi-Albert scale-free networks \cite{barabasi1999emergence} and Erd\"os-R\'enyi random graph \cite{erdos1959random} topologies, as shown in Section A of the Supplementary Information.

This correlation is most clearly seen for small values of the intra-layer coupling strength, such as the value $\sigma_1 = 0.01$ considered in Fig.~\ref{fig1}. For larger values of $\sigma_1$, which make inter-layer synchronization possible with a very small number of steps, the correlation is less strong, while no obvious correlation between the targeting sequence and local topological properties are found for very large $\sigma_1$, see again Section A of the Supplementary Information. This might be related to the enhanced contribution of next-nearest neighbors and other relative distant nodes as the coupling strength is increased. Despite its limited range of validity, this correlation is nontheless remarkable, as it is very robust, and quite different from the situation observed in undirected networks, where the topological observable correlating with the targeting sequence is the degree \cite{gutierrez2012targeting}. On the other hand, there is an intriguing parallel between the correlation reported in  the lower panel of Fig.~\ref{fig1} and the fact that, in undirected networks, nodes with a higher dynamic vulnerability are those with less influence from the rest  of the network, followed by those that  have the strongest ability to influence the rest  \cite{gutierrez2011node}. In fact, both aspects of a node position are combined in the influence index in the case of directed or mixed networks.

\section{Controlling ecological networks}

We next apply our method to a model of a trophic web involving 12 species.  This model describes the dynamics of a generic trophic web, including  several categories of consumers such as top predators ($P_2$, and $P_3$), mesopredators ($M_1$ and $M_2$), several large herbivores  (from $H_1$ to $H_4$), small herbivores ($J_1$ and $J_2$) and also intermediate  omnivourous consumers ($P_1$ and $H_6$) which, in the real world, may also rely on predation and scavenging \cite{focardi2017kleptoparasitism} (cf. Section B of the Supplementary Information for a full description of the model). The  model represents a simplified food web inspired by holarctic ecosystems (see Ref. \cite{focardi2017kleptoparasitism}, and references therein).
Controlling such a trophic web by means of only pinning a limited number of species, and/or implementing desired control policies for specific populations, are tasks of great societal relevance. As a matter of fact, there are many situations where wildlife agencies aim to control populations in order to reduce crop riding, depredation, as well as to control transmissible disease, to reduce extinction risks, or mitigate conflicts among stakeholders (e.g.\! conservationists, farmers, hunters).  

The trophic web is viewed as a network where the species are the different nodes, and the links stand for the interactions among them. From the point of view of network control, this is a challenging model: each node (species) evolves autonomously following different population dynamics, the links (inter-species interactions) are also species-dependent and vary widely both in number, character (some are directed, some are undirected, and they are assigned different strengths) and in the mathematical form of the couplings, which are usually nonlinear. While the details of model are described in Section B of the Supplementary Information, we here briefly summarize its salient qualitative features. Each of the 12 species is described by a scalar that measures the population density at a given time. The coupling between species is given by nonlinear predator-prey response functions and competition-for-resource terms, which are proportional to products of the populations of the competing species.  Moreover, there are logistic growth terms for each of the herbivores. A key feature of this model is the periodic nature of masting, which represents the quasi periodic production of forest fruits, such as acorns. Here masting acts as a forcing agent on the growth rate of one of the populations. The forcing makes the dynamics chaotic, with a (numerically calculated) maximum Lyapunov exponent $\Lambda_\text{max} \simeq 0.0014$.  A representative sample of the highly irregular oscillations of the populations is shown in Section B of the Supplementary Information.

In order to apply the pinning procedure we need to construct a copy of the trophic web from which to establish unidirectional links to the original web. Below we clarify how this can be practically achieved in real ecosystems by monitoring populations along time, but for the time being we assume this to be a feasible task. We then apply pinning actions sequentially until, when a sufficiently high number of pinning actions have been established, the slave layer (the ecosystem of interest) follows the dynamics of the master system. As in the previous section, the key information is contained in the sequence of pinned nodes, as this reveals which are the species whose population one must preserve or modify in order to maintain a desired dynamics or disrupt an undesired one. In actual management, wildlife agencies are often requested to purchase action plans  for removing or reintroducing individuals, to increase recruitment or reduce natural mortality by supplementary feeding, or to modify to some extent the natural dynamics of the system. This makes sense if the action provides long-lasting results, meaning that the ecosystem would attain a new equilibrium.

\begin{figure}[t!]
\hspace{-0.5cm}\includegraphics[scale=0.45]{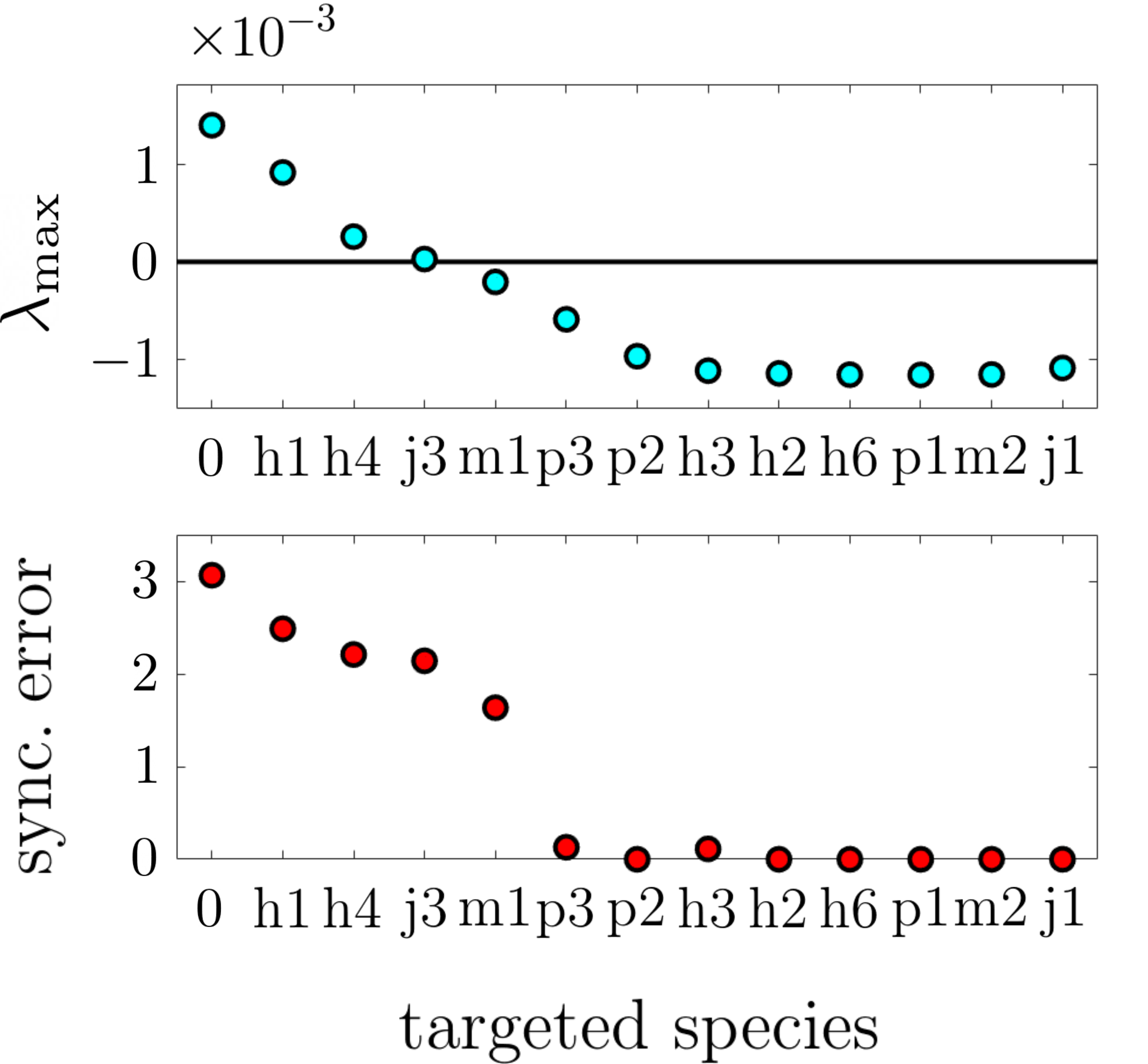}
\caption{ {\sf \bf Controlling the dynamics of a trophic web comprising 12 species with inter-layer coupling $\sigma_2 = 0.005$.}
(Top). Maximum Lyapunov exponent $\lambda_\text{max}$ as a function of the targeting step (codes indicating the species targeted at each step are described in Section B of the Supplementary Information).  (Bottom)  Synchronization error as a function of the targeting step. A 4th-order Runge-Kutta method with a step of 0.01 time units has been employed for the numerical integration of the systems of $12$ ordinary differential equations corresponding to each layer.}\label{fig2}
\end{figure}

In the top panel of Fig.~\ref{fig2}, we show the maximum Lyapunov exponent $\lambda_\text{max}$ as a function of the targeting step, with labels in the horizontal axis indicating the targeted species at each step (see Section B of the Supplementary Information for a detailed description of these codes). It appears that here it is enough to target two species of herbivores, such as deer ($H_1$ and $H_4$), a small herbivore, such as a species of hare ($J_3$), and a mesopredator, such as jackals or foxes ($M_1$), to control the network, to which it is probably necessary to add a control on one large predator, such as the wolf or the coguar if  allowed by laws, $P_3$. Analogous results in terms of the synchronization error are reported in the bottom panel of Fig.~\ref{fig2}, which shows how the synchronization error becomes zero when $\lambda_\text{max}$ becomes sufficiently negative. After pinning four species, $\lambda_\textrm{max}$ is only midly negative and, given the relatively high value of the synchronization error, there appear to be regions in phase space where trajectories diverge despite the fact that the phase space averaging given by the Lyapunov exponent shows that the trend is overall converging. After the fifth pinning (i.e.\! the large predator), both measures unmistakably show that inter-layer synchronization has been achieved. If the sequence of species to be pinned is instead chosen at random, approximately twice as many species have to be pinned to achieve inter-layer synchronization for the same parameter values. The inter-layer parameter is here chosen to be  $\sigma_2 = 0.005$, as considerably larger values lead to instabilities in the dynamics. In this case the intra-layer coupling  is not a free parameter that one can modify at will as in the network of R\"ossler oscillators of the previous section ---in fact, it is determined by the different parameters of the trophic web model and varies from link to link (see Section B of the Supplementary Information for more information).  The length of the sequence of the species needed to achieve synchronization is largely $\sigma_2$ dependent, but the ranking is robust across a range of $\sigma_2$.

In fact, the possibility of imposing on a system a given dynamics compatible with the equations of motion from an initial time onwards may not be always realistic. If the master layer were an exact physical replica of the original system (as could approximately be achieved in networks of nonlinear oscillating circuits or other technological systems) or a faithful simulation of its dynamics, one could hope to achieve that. In most cases, however, one can only expect to obtain (finite) recorded segments of the activity of the system in the form of a time series of some of its observables. Fortunately such a recording periodically repeated may suffice to maintain the system evolution within certain desired region of phase space. This is certainly true in the case of a trophic web, so we next illustrate with our model how the pinning method can be based on a short segment of recorded activity.

\begin{figure}[t!]
\hspace{-0.5cm}\includegraphics[scale=0.33]{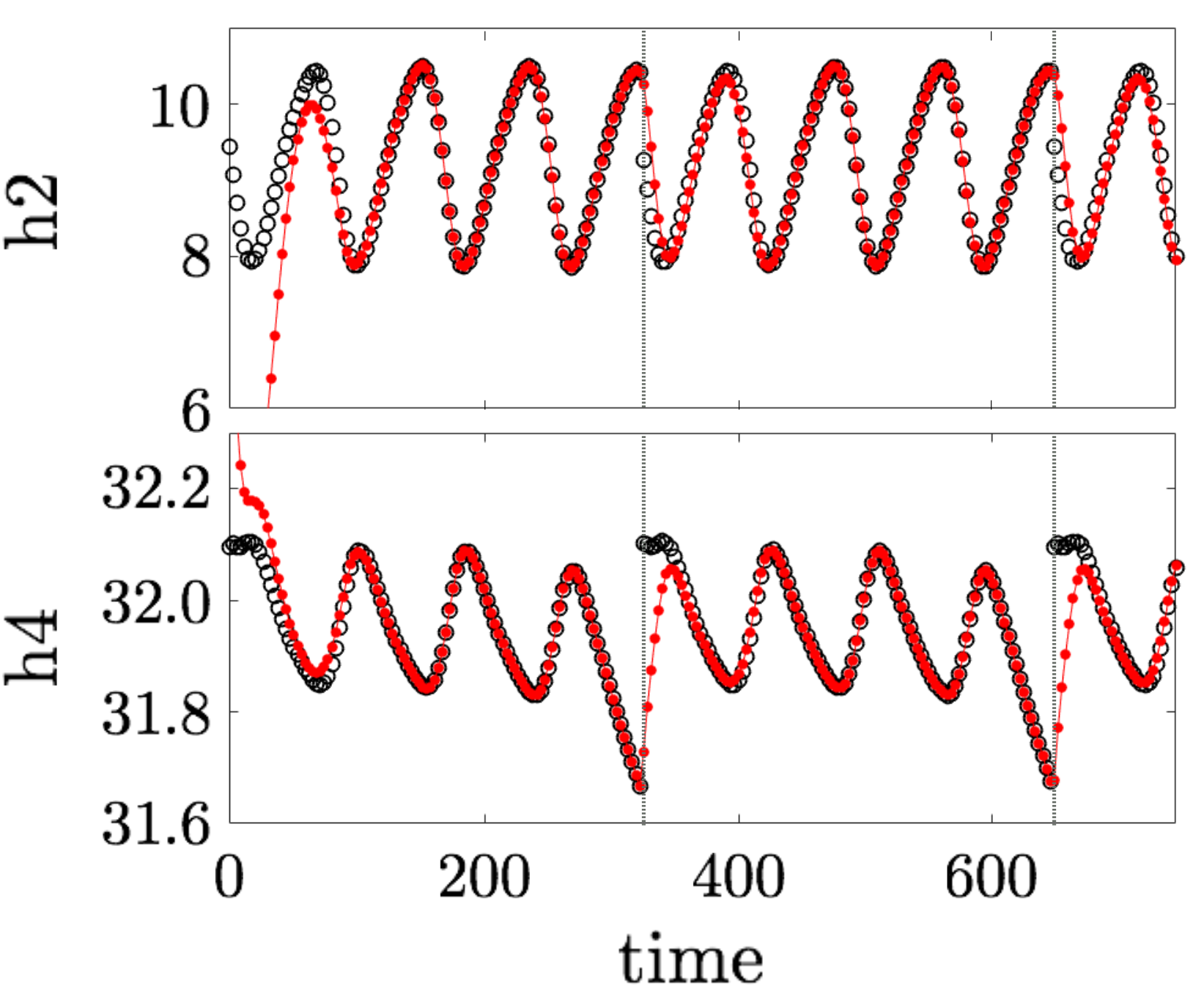}
\caption{ {\sf \bf Evolution of two species of the trophic web (red dotted line) pinned to a periodically repeated segment of recorded population dynamics (black discs).}
(Top) Population of one species of deer as a function of time in the recording and under pinning of 5 species using the recording as master dynamics. 
(Bottom) Population of another species of deer under the same conditions.}\label{fig3}
\end{figure}

To do so, we simulate our trophic web system over a time window of tens of thousands of units and record the species populations every $5$ time units. From this time series (i.e. our recording of ``observational'' data), we choose a time window of 325 time units where the populations oscillate relatively regularly within certain bounds that are of course species-dependent.  Assuming these are the bounds that, for instance, on one side allow the conservation of a given species, but on the other reduce the amount of economic damage to crops, we take this to be our desired dynamics. Our master layer is simply this segment of recorded activity periodically repeated (i.e.\! when we come to the last sample, we start again from the first one, and so on), which we impose on the system by pinning a sufficiently high number of species (we choose this number to be $5$, following the results in Fig.~\ref{fig2}, which also determine the species chosen for the pinning actions). The results are shown in Fig.~\ref{fig3} for two species of the master layer (see black discs), which is just the periodically repeated time window of the recording, and the corresponding species in the slave layer (see red dotted line), which is the actual ecosystem in our model. We see how the pinning rapidly brings the slave system into the desired dynamical regime, despite the fact that it has started from an initial condition which is quite far from it. What we illustrate here for just two species for the sake of brevity, is similarly observed for the remaning ones. While the periodic repetition of the recorded segment introduces some discontinuities in the dynamics, the slave network does not take long (relatively speaking) to follow the master dynamics. In fact the length of the period where the trajectories are visibly different at each start of the cycle is related to the (inverse of the) Lyapunov exponent displayed in Fig.~\ref{fig2}, and in general is expected to become smaller as more species are pinned (at least before the exponent saturates, as happens in the results shown in that figure eventually).

To conclude this section, we should mention that the pinning strategy is also expected to work when not all of the nodes of the network are included in the time series (for instance, when only a subset of the interacting populations are tracked). See the relevant discussion in Ref.~\cite{gutierrez2012targeting}, which is also applicable in the present setting.

\section{Discussion}

We have proposed a quite general procedure for controlling the dynamics of complex networks of nonlinear dynamical units that are coupled in a nonlinear fashion, and possibly through mathematically different coupling functions, according to network schemes that may include unidirectional, bidirectional and weighted links. The method, which was first proposed in a much more restrictive setting \cite{gutierrez2012targeting}, is based on the establishment of unidirectional inter-layer couplings (pinning actions) that reach the nodes of the original networked system from their counterparts in the other layer, which is an identical copy of the system, with the aim of imposing the dynamics of the copy on the network under consideration. While invoking a two-layer structure might sound quite remote from any practical application, in fact the copy can be simply a set of experimental data characterizing the state of the system across time, which may well be finite (in fact quite short) and even contain information on only some of the nodes.

We first illustrate the method on a network of nonlinearly-coupled (chaotic) R\"ossler oscillators coupled through a network topology that includes both directed and undirected links. The sequence of pinned nodes that is found to bring the system closer to the desired dynamics at each step shows a remarkable correlation with a ranking of the network nodes in terms of their influence index $k_\text{out}/k_\text{in}$. This index measures the ability of a node to influence other nodes (as given by the, possibly weighted, out-degree $k_\text{out}$) normalized by the influence other nodes have on it (as given by the in-degree  $k_\text{in}$).

We then move on to a study of a trophic web model containing 12 species inspired by European and North-American ecosystems. This case is much more challenging, as different nodes evolve according to different dynamic rules, and are coupled via nonlinear mathematical functions that depend on the pair of species involved. The method is shown to be perfectly applicable on such systems, and has the potential to yield valuable information on which species are key in maintaining a given ecosystem dynamics. Moreover, we illustrate the method by using as desired dynamics a short segment of activity where the populations evolve within stipulated bounds. Studying correlations between targeting sequences and topological properties of the kind observed in the network of R\"ossler oscillators will require further work in the case of this type of systems, where distinct links represent interactions that are mathematically completely different, and therefore it is hard to give precise operational meanings to measures such as the influence index.

In conclusion, we have presented a versatile method for steering networks toward desired dynamics. This method has shown to be valuable for unveiling correlations between node controllability and topological properties, which provide theoretical insight into the interplay of structure and function in complex systems. Most importantly, the method is of practical value for the control of systems that do not satisfy the highly idealized requirements of network control methods in the literature, systems whose dynamics may not be even fully understood nor amenable to realistic theoretical modelling. In fact, the main challenge facing the application of our method to environmental management is that the dynamics of ecosystems are always imperfectly known and in many  cases scarcely documented. Thus it is necessary to joint this theoretical approach with adaptive management. Adaptive management is a method of learning by doing. Initially the model used as master shall be very rough, but with subsequent refinements based on management actions and monitoring, the method would improve becoming more and more effective.
Our results open a very new avenue to apply adaptive management to nature conservation in the framework of the Convention of Biological Diversity as summarised by the Malawi Principles for the management of whole ecosystems.

\begin{acknowledgments}
The authors acknowledge the computing resources and technical support provided by CRESCO/ENEAGRID High Performance Computing infrastructure (funded by ENEA, the Italian National Agency for New Technologies, Energy and Sustainable Economic Development and by Italian and European research programmes) and its staff \cite{ponti2014}. R.G. acknowledges financial support from MINECO, Spain (FIS2017-84151-P).
\end{acknowledgments}


\begin{thebibliography}{10}
\expandafter\ifx\csname url\endcsname\relax
  \def\url#1{\texttt{#1}}\fi
\expandafter\ifx\csname urlprefix\endcsname\relax\def\urlprefix{URL }\fi
\providecommand{\bibinfo}[2]{#2}
\providecommand{\eprint}[2][]{\url{#2}}

\bibitem{wang2002pinning}
\bibinfo{author}{Wang, X.~F.} \& \bibinfo{author}{Chen, G.}
\newblock \bibinfo{title}{Pinning control of scale-free dynamical networks}.
\newblock \emph{\bibinfo{journal}{Physica A: Statistical Mechanics and its
  Applications}} \textbf{\bibinfo{volume}{310}}, \bibinfo{pages}{521--531}
  (\bibinfo{year}{2002}).

\bibitem{wang2002synchronization}
\bibinfo{author}{Wang, X.~F.} \& \bibinfo{author}{Chen, G.}
\newblock \bibinfo{title}{Synchronization in scale-free dynamical networks:
  robustness and fragility}.
\newblock \emph{\bibinfo{journal}{IEEE Transactions on Circuits and Systems I:
  Fundamental Theory and Applications}} \textbf{\bibinfo{volume}{49}},
  \bibinfo{pages}{54--62} (\bibinfo{year}{2002}).

\bibitem{sorrentino2007controllability}
\bibinfo{author}{Sorrentino, F.}, \bibinfo{author}{Di~Bernardo, M.},
  \bibinfo{author}{Garofalo, F.} \& \bibinfo{author}{Chen, G.}
\newblock \bibinfo{title}{Controllability of complex networks via pinning}.
\newblock \emph{\bibinfo{journal}{Physical Review E}}
  \textbf{\bibinfo{volume}{75}}, \bibinfo{pages}{046103}
  (\bibinfo{year}{2007}).

\bibitem{xiang2007pinning}
\bibinfo{author}{Xiang, L.~Y.}, \bibinfo{author}{Liu, Z.~X.},
  \bibinfo{author}{Chen, Z.~Q.}, \bibinfo{author}{Chen, F.} \&
  \bibinfo{author}{Yuan, Z.~Z.}
\newblock \bibinfo{title}{Pinning control of complex dynamical networks with
  general topology}.
\newblock \emph{\bibinfo{journal}{Physica A: Statistical Mechanics and its
  Applications}} \textbf{\bibinfo{volume}{379}}, \bibinfo{pages}{298--306}
  (\bibinfo{year}{2007}).

\bibitem{boccaletti2014structure}
\bibinfo{author}{Boccaletti, S.} \emph{et~al.}
\newblock \bibinfo{title}{The structure and dynamics of multilayer networks}.
\newblock \emph{\bibinfo{journal}{Physics Reports}}
  \textbf{\bibinfo{volume}{544}}, \bibinfo{pages}{1--122}
  (\bibinfo{year}{2014}).

\bibitem{gutierrez2012targeting}
\bibinfo{author}{Guti{\'e}rrez, R.}, \bibinfo{author}{Sendi\~na Nadal, I.},
  \bibinfo{author}{Zanin, M.}, \bibinfo{author}{Papo, D.} \&
  \bibinfo{author}{Boccaletti, S.}
\newblock \bibinfo{title}{Targeting the dynamics of complex networks}.
\newblock \emph{\bibinfo{journal}{Scientific Reports}}
  \textbf{\bibinfo{volume}{2}}, \bibinfo{pages}{396} (\bibinfo{year}{2012}).

\bibitem{pecora1998master}
\bibinfo{author}{Pecora, L.~M.} \& \bibinfo{author}{Carroll, T.~L.}
\newblock \bibinfo{title}{Master stability functions for synchronized coupled
  systems}.
\newblock \emph{\bibinfo{journal}{Physical Review Letters}}
  \textbf{\bibinfo{volume}{80}}, \bibinfo{pages}{2109} (\bibinfo{year}{1998}).

\bibitem{liu2011controllability}
\bibinfo{author}{Liu, Y.-Y.}, \bibinfo{author}{Slotine, J.-J.} \&
  \bibinfo{author}{Barab{\'a}si, A.-L.}
\newblock \bibinfo{title}{Controllability of complex networks}.
\newblock \emph{\bibinfo{journal}{Nature}} \textbf{\bibinfo{volume}{473}},
  \bibinfo{pages}{167} (\bibinfo{year}{2011}).

\bibitem{posfai2013effect}
\bibinfo{author}{P{\'o}sfai, M.}, \bibinfo{author}{Liu, Y.-Y.},
  \bibinfo{author}{Slotine, J.-J.} \& \bibinfo{author}{Barab{\'a}si, A.-L.}
\newblock \bibinfo{title}{Effect of correlations on network controllability}.
\newblock \emph{\bibinfo{journal}{Scientific Reports}}
  \textbf{\bibinfo{volume}{3}}, \bibinfo{pages}{1067} (\bibinfo{year}{2013}).

\bibitem{jia2013emergence}
\bibinfo{author}{Jia, T.} \emph{et~al.}
\newblock \bibinfo{title}{Emergence of bimodality in controlling complex
  networks}.
\newblock \emph{\bibinfo{journal}{Nature Communications}}
  \textbf{\bibinfo{volume}{4}}, \bibinfo{pages}{1--6} (\bibinfo{year}{2013}).

\bibitem{ruths2014control}
\bibinfo{author}{Ruths, J.} \& \bibinfo{author}{Ruths, D.}
\newblock \bibinfo{title}{Control profiles of complex networks}.
\newblock \emph{\bibinfo{journal}{Science}} \textbf{\bibinfo{volume}{343}},
  \bibinfo{pages}{1373--1376} (\bibinfo{year}{2014}).

\bibitem{nacher2012dominating}
\bibinfo{author}{Nacher, J.~C.} \& \bibinfo{author}{Akutsu, T.}
\newblock \bibinfo{title}{Dominating scale-free networks with variable scaling
  exponent: heterogeneous networks are not difficult to control}.
\newblock \emph{\bibinfo{journal}{New Journal of Physics}}
  \textbf{\bibinfo{volume}{14}}, \bibinfo{pages}{073005}
  (\bibinfo{year}{2012}).

\bibitem{gates2016control}
\bibinfo{author}{Gates, A.~J.} \& \bibinfo{author}{Rocha, L.~M.}
\newblock \bibinfo{title}{Control of complex networks requires both structure
  and dynamics}.
\newblock \emph{\bibinfo{journal}{Scientific Reports}}
  \textbf{\bibinfo{volume}{6}}, \bibinfo{pages}{1--11} (\bibinfo{year}{2016}).

\bibitem{wang2012optimizing}
\bibinfo{author}{Wang, W.-X.}, \bibinfo{author}{Ni, X.}, \bibinfo{author}{Lai,
  Y.-C.} \& \bibinfo{author}{Grebogi, C.}
\newblock \bibinfo{title}{Optimizing controllability of complex networks by
  minimum structural perturbations}.
\newblock \emph{\bibinfo{journal}{Physical Review E}}
  \textbf{\bibinfo{volume}{85}}, \bibinfo{pages}{026115}
  (\bibinfo{year}{2012}).

\bibitem{yuan2013exact}
\bibinfo{author}{Yuan, Z.}, \bibinfo{author}{Zhao, C.}, \bibinfo{author}{Di,
  Z.}, \bibinfo{author}{Wang, W.-X.} \& \bibinfo{author}{Lai, Y.-C.}
\newblock \bibinfo{title}{Exact controllability of complex networks}.
\newblock \emph{\bibinfo{journal}{Nature Communications}}
  \textbf{\bibinfo{volume}{4}}, \bibinfo{pages}{1--9} (\bibinfo{year}{2013}).

\bibitem{gao2014target}
\bibinfo{author}{Gao, J.}, \bibinfo{author}{Liu, Y.-Y.},
  \bibinfo{author}{D'souza, R.~M.} \& \bibinfo{author}{Barab{\'a}si, A.-L.}
\newblock \bibinfo{title}{Target control of complex networks}.
\newblock \emph{\bibinfo{journal}{Nature Communications}}
  \textbf{\bibinfo{volume}{5}}, \bibinfo{pages}{1--8} (\bibinfo{year}{2014}).

\bibitem{cornelius2013realistic}
\bibinfo{author}{Cornelius, S.~P.}, \bibinfo{author}{Kath, W.~L.} \&
  \bibinfo{author}{Motter, A.~E.}
\newblock \bibinfo{title}{Realistic control of network dynamics}.
\newblock \emph{\bibinfo{journal}{Nature Communications}}
  \textbf{\bibinfo{volume}{4}}, \bibinfo{pages}{1--9} (\bibinfo{year}{2013}).

\bibitem{zanudo2017structure}
\bibinfo{author}{Za{\~n}udo, J. G. T.}, \bibinfo{author}{Yang, G.} \&
  \bibinfo{author}{Albert, R.}
\newblock \bibinfo{title}{Structure-based control of complex networks with
  nonlinear dynamics}.
\newblock \emph{\bibinfo{journal}{Proceedings of the National Academy of
  Sciences}} \textbf{\bibinfo{volume}{114}}, \bibinfo{pages}{7234--7239}
  (\bibinfo{year}{2017}).

\bibitem{holling1966functional}
\bibinfo{author}{Holling, C.~S.}
\newblock \bibinfo{title}{The functional response of invertebrate predators to
  prey density}.
\newblock \emph{\bibinfo{journal}{The Memoirs of the Entomological Society of
  Canada}} \textbf{\bibinfo{volume}{98}}, \bibinfo{pages}{5--86}
  (\bibinfo{year}{1966}).

\bibitem{sevilla2016inter}
\bibinfo{author}{Sevilla-Escoboza, R.} \emph{et~al.}
\newblock \bibinfo{title}{Inter-layer synchronization in multiplex networks of
  identical layers}.
\newblock \emph{\bibinfo{journal}{Chaos: An Interdisciplinary Journal of
  Nonlinear Science}} \textbf{\bibinfo{volume}{26}}, \bibinfo{pages}{065304}
  (\bibinfo{year}{2016}).

\bibitem{leyva2017inter}
\bibinfo{author}{Leyva, I.} \emph{et~al.}
\newblock \bibinfo{title}{Inter-layer synchronization in non-identical
  multi-layer networks}.
\newblock \emph{\bibinfo{journal}{Scientific Reports}}
  \textbf{\bibinfo{volume}{7}}, \bibinfo{pages}{45475} (\bibinfo{year}{2017}).

\bibitem{bender1978asymptotic}
\bibinfo{author}{Bender, E.~A.} \& \bibinfo{author}{Canfield, E.~R.}
\newblock \bibinfo{title}{The asymptotic number of labeled graphs with given
  degree sequences}.
\newblock \emph{\bibinfo{journal}{Journal of Combinatorial Theory, Series A}}
  \textbf{\bibinfo{volume}{24}}, \bibinfo{pages}{296--307}
  (\bibinfo{year}{1978}).

\bibitem{liu2008phase}
\bibinfo{author}{Liu, Y.}, \bibinfo{author}{Bi, Q.-S.} \&
  \bibinfo{author}{Chen, Y.-S.}
\newblock \bibinfo{title}{Phase synchronization between nonlinearly coupled
  {R}{\"o}ssler systems}.
\newblock \emph{\bibinfo{journal}{Applied Mathematics and Mechanics}}
  \textbf{\bibinfo{volume}{29}}, \bibinfo{pages}{697} (\bibinfo{year}{2008}).

\bibitem{perko2013differential}
\bibinfo{author}{Perko, L.}
\newblock \emph{\bibinfo{title}{Differential Equations and Dynamical Systems}},
  (\bibinfo{publisher}{Springer-Verlag, New York}, \bibinfo{year}{2001}).

\bibitem{benettin1980lyapunova}
\bibinfo{author}{Benettin, G.}, \bibinfo{author}{Galgani, L.},
  \bibinfo{author}{Giorgilli, A.} \& \bibinfo{author}{Strelcyn, J.-M.}
\newblock \bibinfo{title}{Lyapunov {C}haracteristic {E}xponents for smooth
  dynamical systems and for {H}amiltonian systems; a method for computing all
  of them. part 1: Theory}.
\newblock \emph{\bibinfo{journal}{Meccanica}} \textbf{\bibinfo{volume}{15}},
  \bibinfo{pages}{9--20} (\bibinfo{year}{1980}).

\bibitem{benettin1980lyapunovb}
\bibinfo{author}{Benettin, G.}, \bibinfo{author}{Galgani, L.},
  \bibinfo{author}{Giorgilli, A.} \& \bibinfo{author}{Strelcyn, J.-M.}
\newblock \bibinfo{title}{Lyapunov {C}haracteristic {E}xponents for smooth
  dynamical systems and for {H}amiltonian systems; a method for computing all
  of them. part 2: Numerical application}.
\newblock \emph{\bibinfo{journal}{Meccanica}} \textbf{\bibinfo{volume}{15}},
  \bibinfo{pages}{21--30} (\bibinfo{year}{1980}).

\bibitem{barabasi1999emergence}
\bibinfo{author}{Barab{\'a}si, A.-L.} \& \bibinfo{author}{Albert, R.}
\newblock \bibinfo{title}{Emergence of scaling in random networks}.
\newblock \emph{\bibinfo{journal}{Science}} \textbf{\bibinfo{volume}{286}},
  \bibinfo{pages}{509--512} (\bibinfo{year}{1999}).

\bibitem{erdos1959random}
\bibinfo{author}{Erd{\"o}s, P.} \& \bibinfo{author}{R{\'e}nyi, A.}
\newblock \bibinfo{title}{On random graphs}.
\newblock \emph{\bibinfo{journal}{Publicationes Mathematicae Debrecen}}
  \textbf{\bibinfo{volume}{6}}, \bibinfo{pages}{290--297}
  (\bibinfo{year}{1959}).

\bibitem{gutierrez2011node}
\bibinfo{author}{Guti{\'e}rrez, R.}, \bibinfo{author}{del Pozo, F.} \&
  \bibinfo{author}{Boccaletti, S.}
\newblock \bibinfo{title}{Node vulnerability under finite perturbations in
  complex networks}.
\newblock \emph{\bibinfo{journal}{PLOS ONE}} \textbf{\bibinfo{volume}{6}},
  \bibinfo{pages}{e20236} (\bibinfo{year}{2011}).

\bibitem{focardi2017kleptoparasitism}
\bibinfo{author}{Focardi, S.}, \bibinfo{author}{Materassi, M.},
  \bibinfo{author}{Innocenti, G.} \& \bibinfo{author}{Berzi, D.}
\newblock \bibinfo{title}{Kleptoparasitism and scavenging can stabilize
  ecosystem dynamics}.
\newblock \emph{\bibinfo{journal}{The American Naturalist}}
  \textbf{\bibinfo{volume}{190}}, \bibinfo{pages}{398--409}
  (\bibinfo{year}{2017}).

\bibitem{ponti2014}
\bibinfo{author}{Ponti, G.} \emph{et~al.}
\newblock \bibinfo{title}{The role of medium size facilities in the {HPC}
  ecosystem: the case of the new {CRESCO4} cluster integrated in the {ENEAGRID}
  infrastructure}.
\newblock \emph{\bibinfo{journal}{Proceedings of the 2014 International
  Conference on High Performance Computing and Simulation, HPCS 2014}}
  \bibinfo{pages}{6903807, 1030} (\bibinfo{year}{2014}).

\bibitem{materassi2017kleptoparasitism}
\bibinfo{author}{Materassi, M.}, \bibinfo{author}{Innocenti, G.},
  \bibinfo{author}{Berzi, D.} \& \bibinfo{author}{Focardi, S.}
\newblock \bibinfo{title}{Kleptoparasitism and complexity in a multi-trophic
  web}.
\newblock \emph{\bibinfo{journal}{Ecological Complexity}}
  \textbf{\bibinfo{volume}{29}}, \bibinfo{pages}{49--60}
  (\bibinfo{year}{2017}).

\end{thebibliography}

\begin{thebibliography}{10}
\expandafter\ifx\csname url\endcsname\relax
  \def\url#1{\texttt{#1}}\fi
\expandafter\ifx\csname urlprefix\endcsname\relax\def\urlprefix{URL }\fi
\providecommand{\bibinfo}[2]{#2}
\providecommand{\eprint}[2][]{\url{#2}}

\bibitem{materassi2017kleptoparasitism}
\bibinfo{author}{Materassi, M.}, \bibinfo{author}{Innocenti, G.},
  \bibinfo{author}{Berzi, D.} \& \bibinfo{author}{Focardi, S.}
\newblock \bibinfo{title}{Kleptoparasitism and complexity in a multi-trophic
  web}.
\newblock \emph{\bibinfo{journal}{Ecological Complexity}}
  \textbf{\bibinfo{volume}{29}}, \bibinfo{pages}{49--60}
  (\bibinfo{year}{2017}).

\bibitem{benettin1980lyapunovb}
\bibinfo{author}{Benettin, G.}, \bibinfo{author}{Galgani, L.},
  \bibinfo{author}{Giorgilli, A.} \& \bibinfo{author}{Strelcyn, J.-M.}
\newblock \bibinfo{title}{Lyapunov {C}haracteristic {E}xponents for smooth
  dynamical systems and for {H}amiltonian systems; a method for computing all
  of them. part 2: Numerical application}.
\newblock \emph{\bibinfo{journal}{Meccanica}} \textbf{\bibinfo{volume}{15}},
  \bibinfo{pages}{21--30} (\bibinfo{year}{1980}).



\end{thebibliography}

 \bigskip

\begin{center}
{\Large SUPPLEMENTARY INFORMATION:\\\vspace{0.2cm} Steering complex networks toward desired dynamics}
\end{center}

\renewcommand\thesection{\Alph{section}}
\setcounter{section}{0}

\renewcommand{\thefigure}{A\arabic{figure}}
\renewcommand\theequation{A\arabic{equation}}

\setcounter{figure}{0}
\setcounter{equation}{0}

\section{Correlations of the targeting sequence and the influence index for other network topologies}

\begin{figure}[h!]
\hspace{-0.5cm}\includegraphics[scale=0.30]{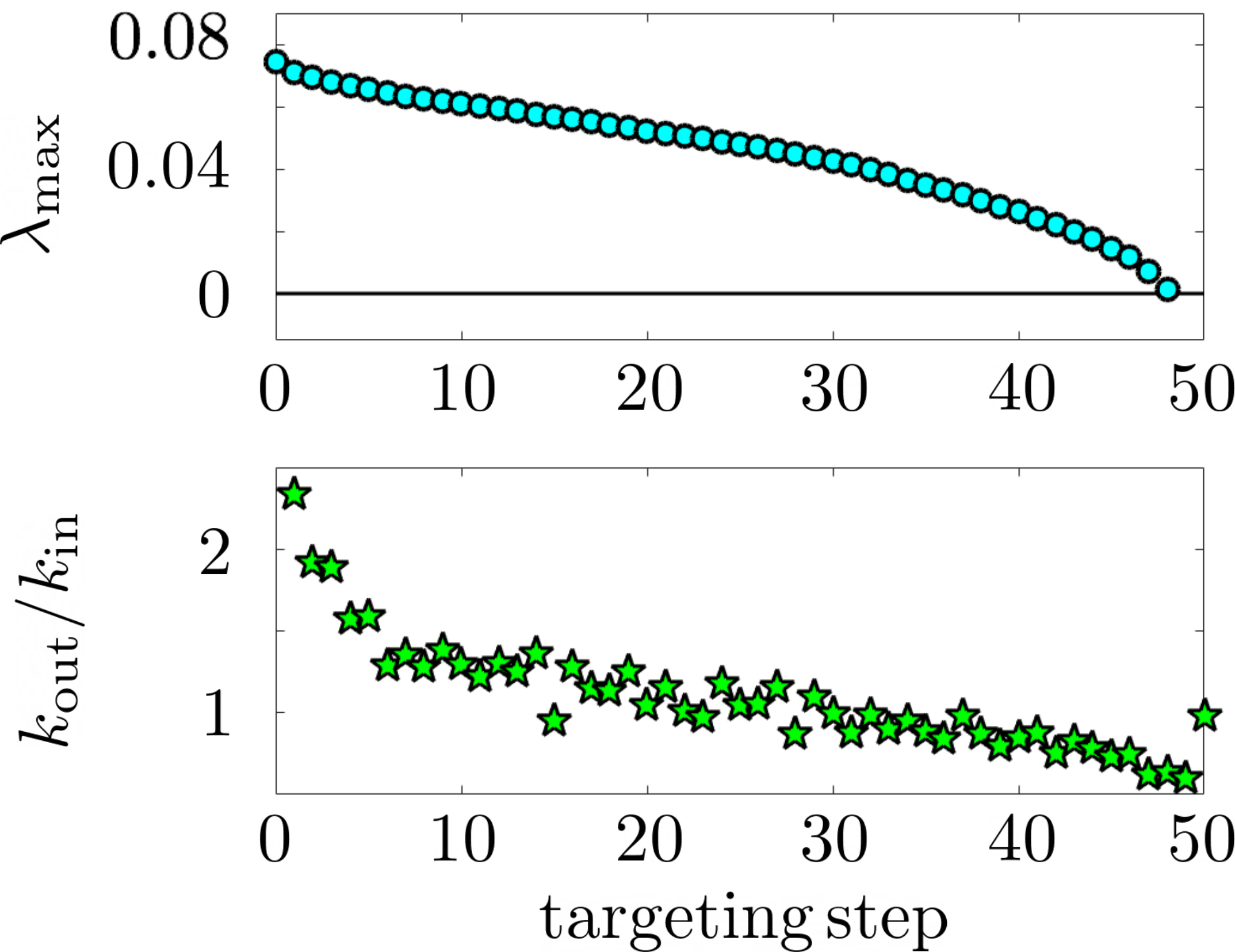}
\caption{ {\sf \bf Controlling the dynamics of a mixed  Erd\"os-R\'enyi random graph of $N=50$ nonlinearly-coupled R\"ossler oscillators with intra-layer coupling $\sigma_1 = 0.01$ and inter-layer coupling $\sigma_2 = 1$.}
(Top). Maximum Lyapunov exponent $\lambda_\text{max}$ as a function of the targeting step.  (Bottom) Influence index $k_\text{out}/k_\text{in}$ of the node that is pinned at each targeting step.}\label{figa22}
\end{figure}

\begin{figure}[h!]
\hspace{-0.5cm}\includegraphics[scale=0.30]{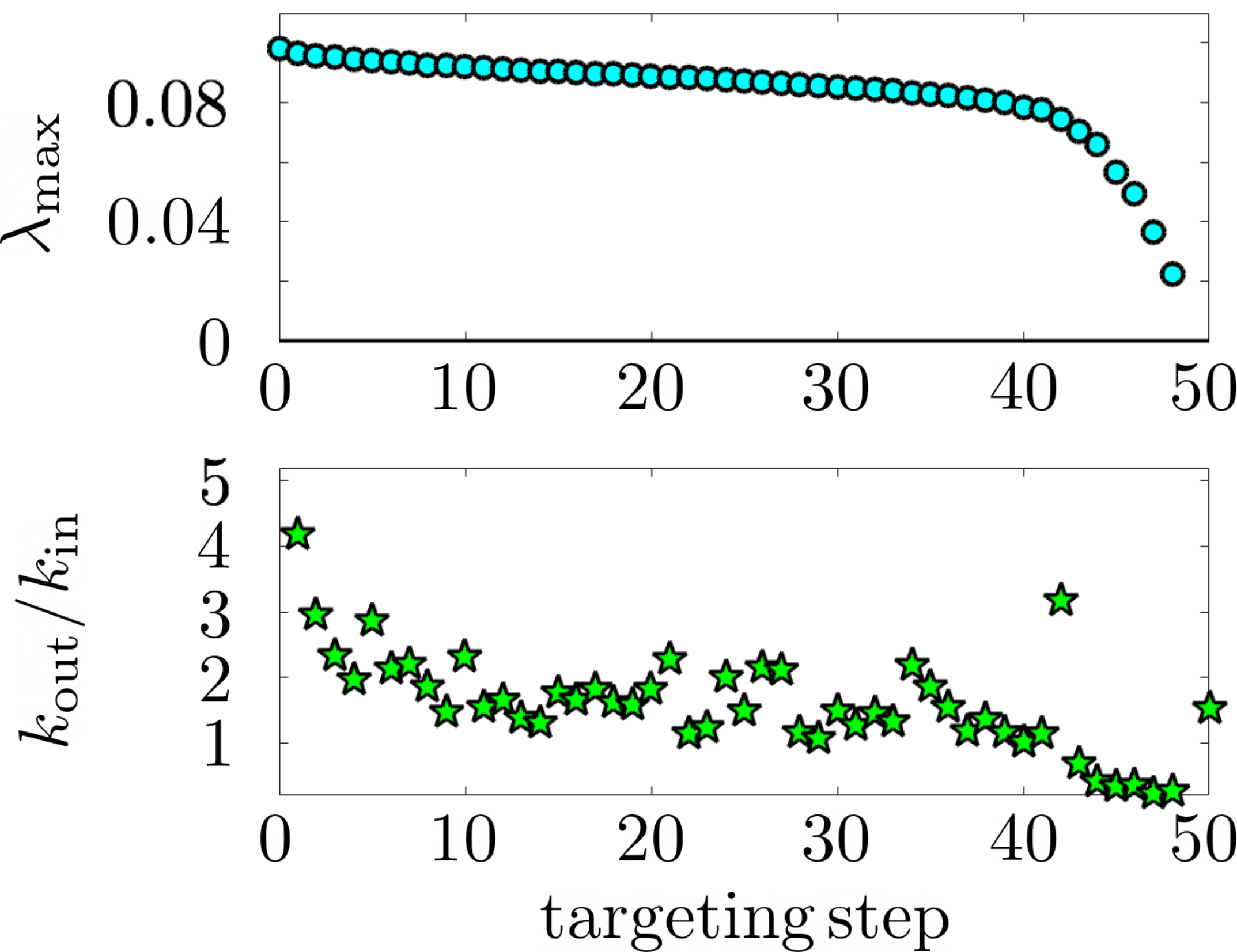}
\caption{ {\sf \bf Controlling the dynamics of a mixed Barab\'asi-Albert scale-free network of $N=50$ nonlinearly-coupled R\"ossler oscillators with intra-layer coupling $\sigma_1 = 0.01$ and inter-layer coupling $\sigma_2 = 1$.}
(Top). Maximum Lyapunov exponent $\lambda_\text{max}$ as a function of the targeting step.  (Bottom) Influence index $k_\text{out}/k_\text{in}$ of the node that is pinned at each targeting step.}\label{figa24}
\end{figure}

As briefly mentioned in the main text, the targeting sequence is correlated with the influence index $k_\text{out}/k_\text{in}$ in the networks with uniform degree distributions considered in Fig.
\!1. And this is also the case with other networks topologies, such as   Erd\"os-R\'enyi random graphs and  Barab\'asi-Albert scale-free networks. Results analogous to those of Fig.\! 1, with the same parameter choices, are shown Fig.~\ref{figa22}, for the  Erd\"os-R\'enyi case, and  Fig.~\ref{figa24}, for the  Barab\'asi-Albert scale-free networks. 

In these figures, the maximum Lyapunov exponent $\lambda_\textrm{max}$ corresponding to the last step of the targeting sequence, which is negative and has a comparatively large absolute value, is not shown in the top panel for visibility reasons. With just 20 network realizations of size  $N=50$, one can clearly see that, except for minor fluctutations, the targeting sequence starts from nodes with high influence index and proceeds toward nodes with a smaller influence in the network.

\begin{figure}[h!]
\hspace{-0.5cm}\includegraphics[scale=0.30]{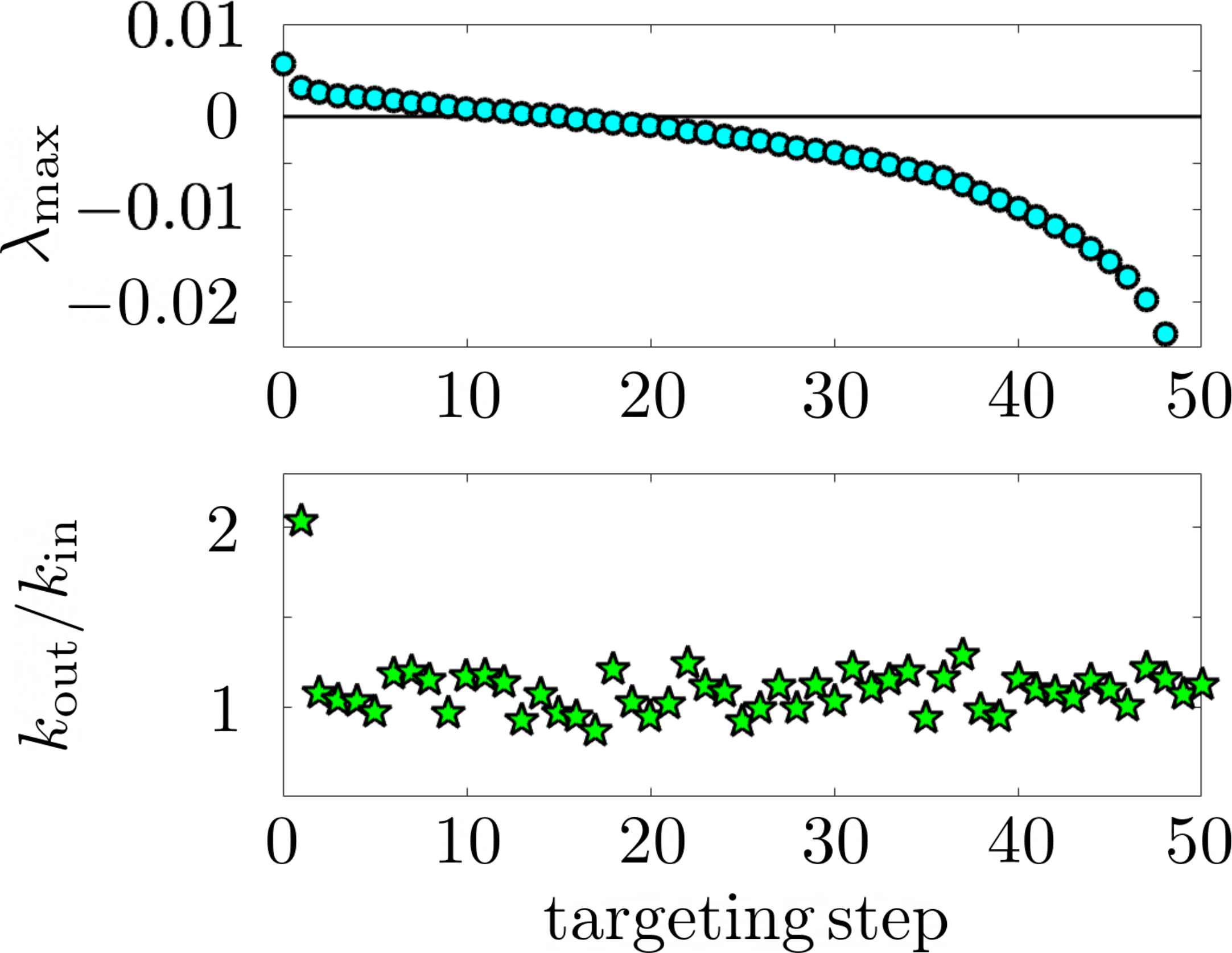}
\caption{ {\sf \bf Controlling the dynamics of a mixed  Erd\"os-R\'enyi random graph of $N=50$ nonlinearly-coupled R\"ossler oscillators with intra-layer coupling $\sigma_1 = 0.05$ and inter-layer coupling $\sigma_2 = 1$.}
(Top). Maximum Lyapunov exponent $\lambda_\text{max}$ as a function of the targeting step.  (Bottom) Influence index $k_\text{out}/k_\text{in}$ of the node that is pinned at each targeting step.}\label{figa23}
\end{figure}

\begin{figure}[h!]
\hspace{-0.5cm}\includegraphics[scale=0.30]{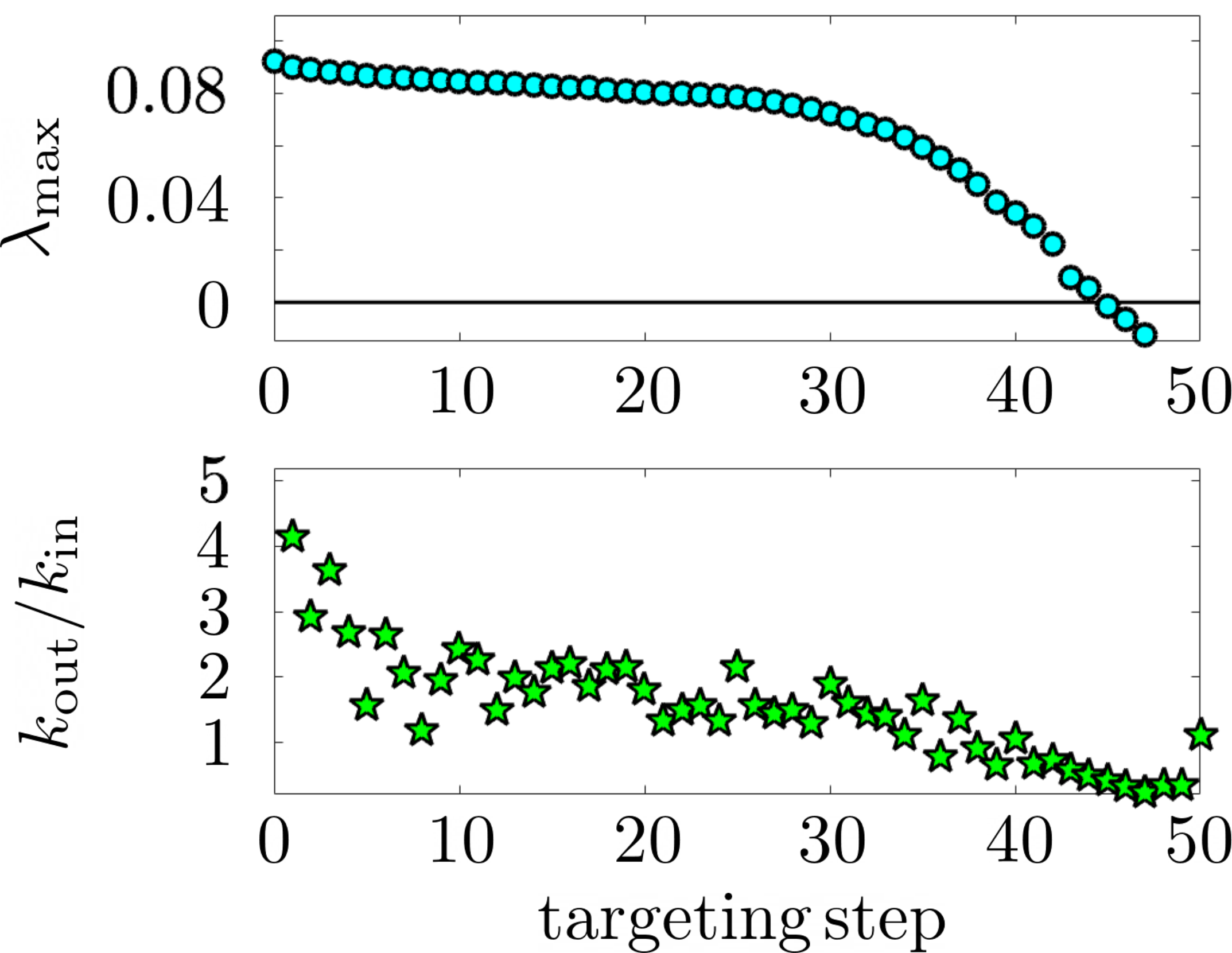}
\caption{ {\sf \bf Controlling the dynamics of a mixed Barab\'asi-Albert scale-free network of $N=50$ nonlinearly-coupled R\"ossler oscillators with intra-layer coupling $\sigma_1 = 0.05$ and inter-layer coupling $\sigma_2 = 1$.}
(Top). Maximum Lyapunov exponent $\lambda_\text{max}$ as a function of the targeting step.  (Bottom) Influence index $k_\text{out}/k_\text{in}$ of the node that is pinned at each targeting step.}\label{figa25}
\end{figure}

While these results, as well as those in Fig.\! 1, correspond to an intra-layer coupling strength  $\sigma_1 = 0.01$ , increasing this value tends to reduce the correlation between targeting sequence and influence index, or even to destroy it altogether. For example,  Fig.~\ref{figa23} shows results analogous to those in  Fig.~\ref{figa22}, for the  Erd\"os-R\'eny case, but with a coupling strength $\sigma_1 = 0.05$, where only the first node in the targeting sequence seems to have a very high influence index $k_\text{out}/k_\text{in}$. Similarly, the network with uniform out-degree $k_\text{out}$ and in-degree $k_\text{in}$ distributions considered in the main text and  Fig.\! 1, seems to show a complete lack of correspondence between the targeting sequence and the influence index ranking when the  coupling strength is $\sigma_1 = 0.05$ (not shown). However, such results may well be related to the fact that for those networks, with such a high intra-layer coupling strength, very few targeting steps are required to achieve inter-layer synchronization. That might also be the reason why in   Fig.~\ref{figa23}, for the  Erd\"os-R\'eny topology, the first node is in fact highly influential, because, by pinning just one node, the network is already very close to inter-layer synchronization. In fact, we do have some evidence that suggests that this could be the case:   Fig.~\ref{figa25} shows results analogous to those in  Fig.~\ref{figa24}, for Barab\'asi-Albert scale-free networks, but with a coupling  strength $\sigma_1 = 0.05$, instead of  $\sigma_1 = 0.01$, where a large number of nodes need to be pinned before inter-layer synchronization is attained. In that case, we do find a clear correlation between the targeting sequence and the influence index ranking.

\renewcommand{\thefigure}{B\arabic{figure}}
\renewcommand\theequation{B\arabic{equation}}

\setcounter{figure}{0}
\setcounter{equation}{0}

\section{Trophic web model}

The considered model mimics the behavior of an actual trophic web of large terrestrial vertebrates. 
It is inspired by holarctic ecosystems found in Asia, Europe and North America. It includes 12 dynamical variables,
representing the densities of 12 species. The variables are:
\begin{itemize}
\item $H_{1}$,  $H_{2}$ ,  $H_{3}$ and  $H_{4}$,  which refer to large herbivores, possibly different species of deer;
\item $H_{6}$, which represents the population of an omnivorous mammal such as the wild boar;
\item $J_{1}$ and $J_3$, which are the number densities of small herbivores, for instance the hare and the beaver;
\item $M_{1}$ and $M_2$, which stand for the populations of two mesopredators, such as the wolverine and the fox;
\item $P_{1}$, which is the number density of a large omnivorous species, in competition with the mesopredators, for instance the bear;
\item $P_{2}$  and $P_{3}$, which refer to large predators, for instance the wolf and the lynx.
\end{itemize}

For simplicity, we assume that herbivores and generalist feeders ($H_1$, $H_2$,
$H_4$, $H_6$, $J_1$, $J_3$, $M_2$ and $P_1$) follow a logistic growth:
\begin{equation}
\dot{X}=R\left(1-\frac{X}{K}\right)X,\label{eq:autotrofa}
\end{equation}
while the mesopredator $M_{1}$ and the predators $P_{2}$ and $P_{3}$
undergo stand-alone equations of exponential decay:
\begin{equation}
Y=-\Delta Y\label{eq:eterotrofa}.
\end{equation}
Notice that in Eqs. (\ref{eq:autotrofa}) and (\ref{eq:eterotrofa}), 
$R$, $K$ and $\Delta$ are all population-independent quantities.
In the absence
of other populations, Eq. (\ref{eq:autotrofa}) gives rise to 
a growth dynamics which asymptotically reaches the value $X=K$; while Eq. (\ref{eq:eterotrofa})
will lead to extinction of the species on a time scale
of the order of $\frac{1}{\Delta}$. This means that populations $H_{1}$,
$H_{2}$, $H_{4}$, $H_{6}$, $J_{1}$, $J_{3}$, $M_{2}$ and $P_{1}$
may be sustained ``by the environment'' in the absence of the other
species, while $M_{1}$, $P_{2}$ and $P_{3}$ will starve if left
alone. In other words, populations
$M_{1}$, $P_{2}$ and $P_{3}$ can live only when the other species are present,
while populations $H_{1}$, $H_{2}$, $H_{4}$,
$H_{6}$, $J_{1}$, $J_{3}$, $M_{2}$ and $P_{1}$ are fed by resources which are
not included in the trophic web model.

As for the growth rate coefficient $R$, this
is a constant quantity for all species except for the omnivorous
 $H_{6}$, for which one has instead
\begin{equation}
R_{H_{6}}\left(t\right)=R_{0H_{6}}+W_{H_{6}}\sin^{2}\left(\dfrac{\omega t}{2}\right).\label{eq:acorn}
\end{equation}
The above Equation accounts for a time-dependent, periodic resource supply from the environment
to the population $H_{6}$ (for instance it may refer to the periodic hyper-production
of acorn, if $H_{6}$ describes wild boars).

The interactions between populations may be
of two kinds: either prey-predator-like, or competitive.

If a population $Y$ preys on the population $X$, this brings a term
$\dot{X}_{Y}^{\mathrm{pred}}$ in the equation for $X$, and a term $\dot{Y}_{X}^{\mathrm{pred}}$
in that for $Y$. The evolution equations describe the case of \emph{satiable predators} \cite{materassi2017kleptoparasitism}, and take the form 
\begin{equation}
\begin{cases}
 & \dot{X}_{Y}^{\mathrm{pred}}=-f_{XY}\left(X\right)Y,\ \dot{Y}_{X}^{\mathrm{pred}}=C_{XY}f_{XY}\left(X\right)Y,\\
\\
 & f_{XY}\left(X\right)=\dfrac{A_{XY}X}{B_{XY}+Y}.
\end{cases}\label{eq:predazione}
\end{equation}
The coefficients $A_{XY}$, $B_{XY}$ and $C_{XY}$ are all positive
constants: if $\dot{X}_{Y}^{\mathrm{pred}}$ is negative,
then $\dot{Y}_{X}^{\mathrm{pred}}$ is larger than zero, as predation
is advantageous for the predator and disadvantageous for the prey.

On the other hand, competitive interactions  lead to two negative terms in the equations of the species $X$ and $Y$:
\begin{equation}
\dot{X}_{Y}^{\mathrm{comp}}=-\alpha_{XY}XY,\ \dot{Y}_{X}^{\mathrm{comp}}=-\alpha_{YX}XY.\label{eq:competizione}
\end{equation}
The two quantities $\alpha_{XY}$ and $\alpha_{YX}$ are both positive,
and not necessarily equal: indeed, competition may be more disadvantageous
for one species than for the other one.

Therefore, the whole trophic web can be described by the following system of differential equations:\\

Equation for $H_{1}$:
\begin{equation}
\begin{array}{c}
\dot{H}_{1}=R_{H_{1}}\left(1-{\displaystyle \frac{H_{1}}{K_{H_{1}}}}\right)H_{1}+\\
\\
-\alpha_{13}H_{1}H_{3}-\alpha_{12}H_{2}H_{1}-\alpha_{14}H_{1}H_{4}-\alpha_{16}H_{1}H_{6}+\\
\\
-f_{H_{1}M_{1}}\left(H_{1}\right)M_{1}-f_{H_{1}M_{2}}\left(H_{1}\right)M_{2}+\\
\\
-f_{H_{1}P_{2}}\left(H_{1}\right)P_{2}-f_{H_{1}P_{3}}\left(H_{1}\right)P_{3}.
\end{array}\label{eq:ODE.H1}
\end{equation}
Equation for $H_{2}$:
\begin{equation}
\begin{array}{c}
\dot{H}_{2}=R_{H_{2}}\left(1-{\displaystyle \frac{H_{2}}{K_{H_{2}}}}\right)H_{2}+\\
\\
-\alpha_{21}H_{1}H_{2}-\alpha_{23}H_{2}H_{3}-\alpha_{24}H_{2}H_{4}-\alpha_{26}H_{2}H_{6}+\\
\\
-f_{H_{2}M_{1}}\left(H_{2}\right)M_{1}-f_{H_{2}M_{2}}\left(H_{2}\right)M_{2}+\\
\\
-f_{H_{2}P_{2}}\left(H_{2}\right)P_{2}-f_{H_{2}P_{3}}\left(H_{2}\right)P_{3}.
\end{array}\label{eq:ODE.H2}
\end{equation}
Equation for $H_{3}$:
\begin{equation}
\begin{array}{c}
\dot{H}_{3}=R_{H_{3}}\left(1-{\displaystyle \frac{H_{3}}{K_{H_{3}}}}\right)H_{3}+\\
\\
-\alpha_{31}H_{1}H_{3}-\alpha_{32}H_{2}H_{3}-\alpha_{34}H_{3}H_{4}-\alpha_{36}H_{3}H_{6}+\\
\\
-f_{H_{3}M_{1}}\left(H_{3}\right)M_{1}-f_{H_{3}M_{2}}\left(H_{3}\right)M_{2}+\\
\\
-f_{H_{3}P_{2}}\left(H_{3}\right)P_{2}-f_{H_{3}P_{3}}\left(H_{3}\right)P_{3}.
\end{array}\label{eq:ODE.H3}
\end{equation}
Equation for $H_{4}$:
\begin{equation}
\begin{array}{c}
\dot{H}_{4}=R_{H_{4}}\left(1-{\displaystyle \frac{H_{4}}{K_{H_{4}}}}\right)H_{4}+\\
\\
-\alpha_{41}H_{1}H_{4}-\alpha_{42}H_{2}H_{4}-\alpha_{43}H_{3}H_{4}-\alpha_{46}H_{4}H_{6}+\\
\\
-f_{H_{4}M_{2}}\left(H_{4}\right)M_{2}+\\
\\
-f_{H_{4}P_{2}}\left(H_{4}\right)P_{2}-f_{H_{4}P_{3}}\left(H_{4}\right)P_{3}.
\end{array}\label{eq:ODE.H4}
\end{equation}
Equations for $H_{6}$:
\begin{equation}
\begin{array}{c}
\dot{H}_{6}=\left[R_{0H_{6}}+W_{H_{6}}\sin^{2}\left(\dfrac{\omega t}{2}\right)\right]\left(1-{\displaystyle \frac{H_{6}}{K_{H_{6}}}}\right)H_{6}+\\
\\
-\alpha_{61}H_{1}H_{6}-\alpha_{62}H_{2}H_{6}-\alpha_{63}H_{3}H_{6}+\\
\\
-f_{H_{6}M_{2}}\left(H_{6}\right)M_{2}+\\
\\
-f_{H_{6}P_{2}}\left(H_{6}\right)P_{2}-f_{H_{6}P_{3}}\left(H_{6}\right)P_{3}
\end{array}\label{eq:ODE.H6.acorn.prod}
\end{equation}

Equations for $J_{1}$:
\begin{equation}
\begin{array}{c}
\dot{J}_{1}=R_{J_{1}}\left(1-{\displaystyle \frac{J_{1}}{K_{J_{1}}}}\right)J_{1}+\\
\\
-f_{J_{1}M_{1}}\left(J_{1}\right)M_{1}-f_{J_{1}M_{2}}\left(J_{1}\right)M_{2}+\\
\\
-f_{J_{1}P_{3}}\left(J_{1}\right)P_{3}.
\end{array}\label{eq:ODE.J1}
\end{equation}
Equations for $J_{3}$:
\begin{equation}
\begin{array}{c}
\dot{J}_{3}=R_{J_{3}}\left(1-{\displaystyle \frac{J_{3}}{K_{J_{3}}}}\right)J_{3}+\\
\\
-f_{J_{3}M_{1}}\left(J_{3}\right)M_{1}-f_{J_{3}M_{2}}\left(J_{3}\right)M_{2}.
\end{array}\label{eq:ODE.J3}
\end{equation}
Equations for $M_{1}$:
\begin{equation}
\begin{array}{c}
\dot{M}_{1}=-\Delta_{M_{1}}M_{1}+\\
\\
-\beta_{11}M_{1}P_{1}-\beta_{12}M_{1}P_{2}-\beta_{13}M_{1}P_{3}+\\
\\
+\left[C_{H_{1}M_{1}}f_{H_{1}M_{1}}\left(H_{1}\right)+C_{H_{2}M_{1}}f_{H_{2}M_{1}}\left(H_{2}\right)+\right.\\
\\
+C_{H_{3}M_{1}}f_{H_{3}M_{1}}\left(H_{3}\right)+\\
\\
\left.+C_{J_{1}M_{1}}f_{J_{1}M_{1}}\left(J_{1}\right)+C_{J_{3}M_{1}}f_{J_{3}M_{1}}\left(J_{3}\right)\right]M_{1}.
\end{array}\label{eq:ODE.M1}
\end{equation}
Equations for $M_{2}$:
\begin{equation}
\begin{array}{c}
\dot{M}_{2}=R_{M_{2}}\left(1-{\displaystyle \frac{M_{2}}{K_{M_{2}}}}\right)M_{2}+\\
\\
-\beta_{21}M_{2}P_{1}-\beta_{22}M_{2}P_{2}-\beta_{23}M_{2}P_{3}+\\
\\
+\left[C_{H_{1}M_{2}}f_{H_{1}M_{2}}\left(H_{1}\right)+C_{H_{2}M_{2}}f_{H_{2}M_{2}}\left(H_{2}\right)+\right.\\
\\
+C_{H_{3}M_{2}}f_{H_{3}M_{2}}\left(H_{3}\right)+C_{H_{4}M_{2}}f_{H_{4}M_{2}}\left(H_{4}\right)+\\
\\
+C_{H_{6}M_{2}}f_{H_{6}M_{2}}\left(H_{6}\right)+\\
\\
\left.+C_{J_{1}M_{2}}f_{J_{1}M_{2}}\left(J_{1}\right)+C_{J_{3}M_{2}}f_{J_{3}M_{2}}\left(J_{3}\right)\right]M_{2}.
\end{array}\label{eq:ODE.M2}
\end{equation}
Equations for $P_{1}$:
\begin{equation}
\begin{array}{c}
\dot{P}_{1}=R_{P_{1}}P_{1}\left(1-{\displaystyle \frac{P_{1}}{K_{P_{1}}}}\right)+\\
\\
-\gamma_{11}P_{1}M_{1}-\gamma_{12}P_{1}M_{2}.
\end{array}\label{eq:ODE.P1.logistic.growth}
\end{equation}
Equations for $P_{2}$:
\begin{equation}
\begin{array}{c}
\dot{P}_{2}=-\Delta_{P_{2}}P_{2}+\\
\\
-\gamma_{21}P_{2}M_{1}-\gamma_{22}P_{2}M_{2}+\\
\\
+\left[C_{H_{1}P_{2}}f_{H_{1}P_{2}}\left(H_{1}\right)+C_{H_{2}P_{2}}f_{H_{2}P_{2}}\left(H_{2}\right)+\right.\\
\\
+C_{H_{3}P_{2}}f_{H_{3}P_{2}}\left(H_{3}\right)+\\
\\
\left.+C_{H_{4}P_{2}}f_{H_{4}P_{2}}\left(H_{4}\right)+C_{H_{6}P_{2}}f_{H_{6}P_{2}}\left(H_{6}\right)\right]P_{2}.
\end{array}\label{eq:ODE.P2}
\end{equation}
Equations for $P_{3}$:
\begin{equation}
\begin{array}{c}
\dot{P}_{3}=-\Delta_{P_{3}}P_{3}+\\
\\
-\gamma_{31}P_{3}M_{1}-\gamma_{32}P_{3}M_{2}+\\
\\
+\left[C_{J_{1}P_{3}}f_{J_{1}P_{3}}\left(J_{1}\right)+C_{H_{1}P_{3}}f_{H_{1}P_{3}}\left(H_{1}\right)+\right.\\
\\
+C_{H_{2}P_{3}}f_{H_{2}P_{3}}\left(H_{2}\right)+C_{H_{3}P_{3}}f_{H_{3}P_{3}}\left(H_{3}\right)+\\
\\
\left.+C_{H_{4}P_{3}}f_{H_{4}P_{3}}\left(H_{4}\right)+C_{H_{6}P_{3}}f_{H_{6}P_{3}}\left(H_{6}\right)\right]P_{3}.
\end{array}\label{eq:ODE.P3}
\end{equation}
For simplicity, the coefficients introduced as $\alpha_{XY}$
and $\alpha_{YX}$ in Eq.(\ref{eq:competizione}) have been renamed (all over the equations) as $\beta$
and $\gamma$, respectively.

The parameters appearing in the model are assigned as follows.\\ 

Parameters for $H_{1}$:
\[
R_{H_{1}} = 0.5, \ K_{H_{1}}=50,
\]
\[
A_{H_{1}M_{1}}=\frac{4}{15},\ B_{H_{1}M_{1}}=20,\ A_{H_{1}M_{2}}=\frac{7}{16},\ B_{H_{1}M_{2}}=20,
\]
\[
A_{H_{1}P_{2}}=\frac{5}{18},\ B_{H_{1}P_{2}}=20,\ A_{H_{1}P_{3}}=\frac{18}{89},\ B_{H_{1}P_{3}}=15,
\]
\[
\alpha_{12}=2\times10^{-4},\ \alpha_{13}=3\times10^{-4},\ \alpha_{14}=5\times10^{-4},\ \alpha_{16}=10^{-5}.
\]
Parameters for $H_{2}$:
\[
R_{H_{2}} = 0.25,\ K_{H_{2}}=15,
\]
\[
A_{H_{2}M_{1}}=0.05,\ B_{H_{2}M_{1}}=8,\ A_{H_{2}M_{2}}=0.075,\ B_{H_{2}M_{2}}=8,
\]
\[
A_{H_{2}P_{2}}=0.1,\ B_{H_{2}P_{2}}=10,\ A_{H_{2}P_{3}}=\frac{1}{34},\ B_{H_{2}P_{3}}=10,
\]
\[
\alpha_{21}=10^{-6},\ \alpha_{23}=3\times10^{-5},\ \alpha_{24}=5\times10^{-4},\ \alpha_{26}=10^{-6}.
\]
Parameters for $H_{3}$:

\[
R_{H_{3}}=0.45,\ K_{H_{3}}=2,
\]
\[
A_{H_{3}M_{1}}=10^{-4},\ B_{H_{3}M_{1}}=1,\ A_{H_{3}M_{2}}=10^{-5},\ B_{H_{3}M_{2}}=1,
\]
\[
A_{H_{3}P_{2}}=0.06,\ B_{H_{3}P_{2}}=1.5,\ A_{H_{3}P_{3}}=\frac{2}{325},\ B_{H_{3}P_{3}}=1.5,
\]
\[
\alpha_{31}=10^{-6},\ \alpha_{32}=10^{-4},\ \alpha_{34}=10^{-6},\ \alpha_{36}=10^{-6}.
\]
Parameters for $H_{4}$ :
\[
R_{H_{4}}=0.3,\ K_{H_{4}}=40,
\]
\[
A_{H_{4}M_{2}}=0.5,\ B_{H_{4}M_{2}}=25,\ A_{H_{4}P_{2}}=\frac{7}{30},\ B_{H_{4}P_{2}}=15,
\]
\[
A_{H_{4}P_{3}}=0.075,\ B_{H_{4}P_{3}}=15,\ \alpha_{41}=10^{-4},\ \alpha_{42}=2\times10^{-4},
\]
\[
\alpha_{43}=10^{-5},\ \alpha_{46}=10^{-4}.
\]
Parameters for $H_{6}$:
\[
R_{H_{6}}=0.8,\, W_{H_{6}} = 5,\, K_{H_{6}}=40,
\]
\[
A_{H_{6}P_{2}}=0.5,\ B_{H_{6}P_{2}}=20,\ A_{H_{6}P_{3}}=0.3,\ B_{H_{6}P_{3}}=15,
\]
\[
\ A_{H_{6}M_{2}}=0,\ B_{H_{6}M_{2}}=20,
\]
\[
\alpha_{61}=10^{-5},\ \alpha_{62}=10^{-5},\ \alpha_{63}=10^{-5}.
\]

Parameters for $J_{1}$:
\[
R_{J_{1}}=0.8,\ K_{J_{1}}=100,
\]
\[
A_{J_{1}M_{1}}=2.2,\ B_{J_{1}M_{1}}=50,\ A_{J_{1}M_{2}}=0.1,\ B_{J_{1}M_{2}}=50,
\]
\[
A_{J_{1}P_{3}}=0.5,\ B_{J_{1}P_{3}}=90.
\]
Parameters for $J_{3}$:
\[
R_{J_{3}}=1.2,\ K_{J_{3}}=15,
\]
\[
A_{J_{3}M_{1}}=1,\ B_{J_{3}M_{1}}=5,\ A_{J_{3}M_{2}}=0.1,\ B_{J_{3}M_{2}}=5.
\]
Parameters for $M_{1}$:
\[
\Delta_{M_{1}}=\frac{1}{15},\ \beta_{11}=10^{-3},\ \beta_{12}=5\times10^{-3},\ \beta_{13}=10^{-4},
\]
\[
C_{H_{1}M_{1}}=0.25,\ C_{H_{2}M_{1}}=0.35,\ C_{H_{3}M_{1}}=0.4,
\]
\[
C_{J_{1}M_{1}}=0.05,\ C_{J_{3}M_{1}}=0.05.
\]
Parameters for $M_{2}$:
\[
R_{M_{2}}=1.2,\ K_{M_{2}}=5,
\]
\[
\beta_{21}=10^{-3},\ \beta_{22}=5\times10^{-3},\ \beta_{23}=10^{-4},
\]
\[
C_{H_{1}M_{2}}=0.3,\ C_{H_{2}M_{2}}=0.35,\ C_{H_{3}M_{2}}=0.35,
\]
\[
C_{H_{4}M_{2}}=0.35,\ C_{H_{6}M_{2}}=0.3,
\]
\[
C_{J_{1}M_{2}}=0.05,\ C_{J_{3}M_{2}}=0.05.
\]

\begin{figure}[h!]
\hspace{-0.5cm}\includegraphics[scale=0.25]{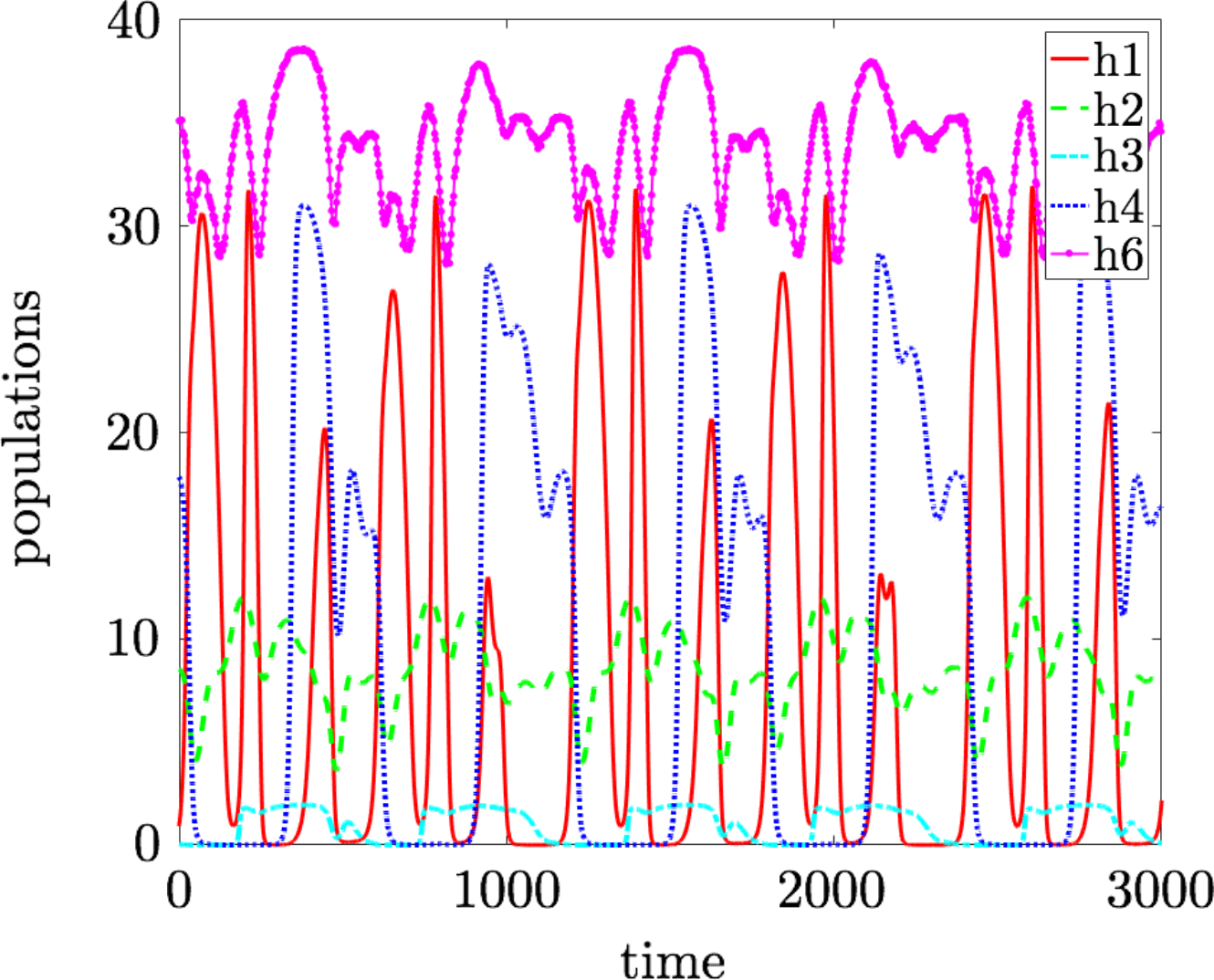}
\caption{Time evolution of the populations of large herbivores (H1-H4) and of the omnivorous mammal (H6). The color code is specified in the legend.}\label{figA1}
\end{figure}

Parameters for $P_{1}$:
\[
R_{P_{1}}=0.25,\ K_{P_{1}}=0.1,
\]
\[
\gamma_{11}=10^{-6},\ \gamma_{12}=10^{-6}.
\]
Parameters for $P_{2}$:
\[
\Delta_{P_{2}}=0.3,\ \gamma_{21}=10^{-6},\ \gamma_{22}=10^{-6},\ C_{H_{1}P_{2}}=0.4,
\]
\[
C_{H_{2}P_{2}}=0.7,\ C_{H_{3}P_{2}}=0.8,\ C_{H_{4}P_{2}}=0.6,\ C_{H_{6}P_{2}}=0.6.
\]
Parameters for $P_{3}$:
\[
\Delta_{P_{3}}=0.1,\ \gamma_{31}=10^{-6},\ \gamma_{32}=10^{-6},
\]
\[
C_{J_{1}P_{3}}=0.05,\ C_{H_{1}P_{3}}=0.3,\ C_{H_{2}P_{3}}=0.3,
\]
\[
C_{H_{3}P_{3}}=0.4,\ C_{H_{4}P_{3}}=0.3,\ C_{H_{6}P_{3}}=0.3.
\]


\begin{figure}[h!]
\hspace{-0.5cm}\includegraphics[scale=0.25]{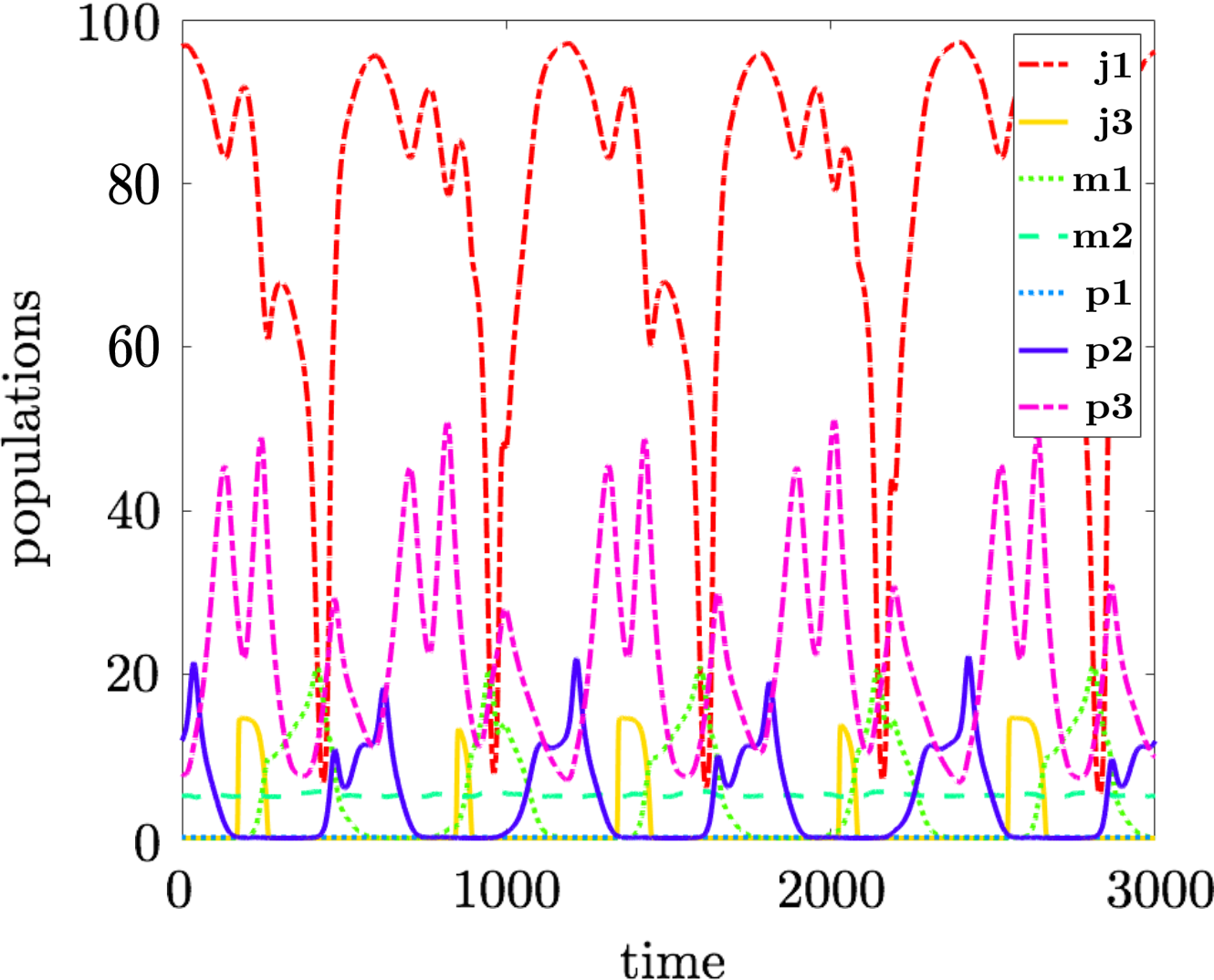}
\caption{Time evolution of the populations of small herbivores (J1 and J3), mesopredators (M1 and M2), a large omnivorous species (P1) and large predators (P2 and P3). The color code is specified in the legend.}\label{figA2}
\end{figure}

With this parameter choice, the twelve species display chaotic oscillations, with a maximum Lyapunov exponent $\Lambda_\text{max} \simeq 0.0014$, which was calculated by use of the method
introduced in Ref. \cite{benettin1980lyapunovb}. In Fig.~\ref{figA1}, irregular oscillations are reported which characterize the evolution of the large herbivors (H1-H4) and of
the omnivorous mammal (H6). The remaining species, including  small herbivores (J1 and J3), mesopredators (M1 and M2), a large omnivorous species (P1) and large predators (P2 and P3), are shown in Fig. \ref{figA2}.

\end{document}